\documentclass[aps,preprint,nofootinbib,eqsecnum]{revtex4}%
\usepackage{amsmath}
\usepackage{graphicx}
\usepackage{amsfonts}
\usepackage{amssymb}%
\providecommand{\U}[1]{\protect\rule{.1in}{.1in}}
\providecommand{\U}[1]{\protect\rule{.1in}{.1in}}
\providecommand{\U}[1]{\protect\rule{.1in}{.1in}}
\providecommand{\U}[1]{\protect\rule{.1in}{.1in}}

\begin{document}

\begin{center}

{\leftline {USC-08/HEP-B1 \hfill arXiv:0802.1947v1 [hep-th]}}

{\vskip0.8cm}

{\Large Dualities among 1T- Field Theories with Spin }

{\Large Emerging from a Unifying 2T- Field Theory}\footnote{This work was
partially supported by the US Department of Energy, grant number
DE-FG03-84ER40168.}

\medskip\textbf{Itzhak Bars and Guillaume Qu\'{e}lin}

\medskip\textsl{Department of Physics and Astronomy, }

\textsl{University of Southern California, Los Angeles, CA 90089-0484, USA.}

\bigskip\textbf{Abstract}
\end{center}

The relation between two time physics (2T-physics) and the ordinary
one time formulation of physics (1T-physics) is similar to the
relation between a 3-dimensional object moving in a room and its
multiple shadows moving on walls when projected from different
perspectives. The multiple shadows as seen by observers stuck on the
wall are analogous to the effects of the 2T-universe as experienced
in ordinary 1T spacetime. In this paper we develop some of the
quantitative aspects of this 2T to 1T relationship in the context of
field theory. We discuss 2T field theory in d+2 dimensions and its
shadows in the form of 1T field theories when the theory contains
Klein-Gordon, Dirac and Yang-Mills fields, such as the Standard
Model of particles and forces. We show that the shadow 1T field
theories must have hidden relations among themselves. These
relations take the form of dualities and hidden spacetime
symmetries. A subset of the shadows are 1T field theories in
different gravitational backgrounds (different space-times) such as
the flat Minkowski spacetime, the Robertson-Walker expanding
universe, AdS$_{d-k}\times$S$^{k},$ and others, including singular
ones. We explicitly construct the duality transformations among this
conformally flat subset, and build the generators of their hidden
SO(d,2) symmetry. The existence of such hidden relations among 1T
field theories, which can be tested by both theory and experiment in
1T-physics, is part of the evidence for the underlying d+2
dimensional spacetime and the unifying 2T-physics structure.

\newpage

\tableofcontents
\newpage

\section{Allegory on the relation between 1T and 2T Physics}

The physical content of 2T-physics and its relation to 1T-physics may be
described with an allegory. The allegory is to consider a 3-dimensional object
moving in a room and the relationships among different shadows of the same
object when projected on 2-dimensional walls by shining light on it from
different perspectives. To observers that live only on the walls (similar to
living only in 3+1 dimensions) the different shadows appear as different
\textquotedblleft beasts\textquotedblright\ (like different 1T-physics
systems). But with hard work, observers on the wall will discover enough
relationships among the shadows to reconstruct the 3 dimensional object.

The allegory above applies because, due to a richer set of gauge symmetry
constraints, 2T-physics in $4+2$ dimensions with 2 times effectively behaves
like 1T-physics in 3+1 dimensions with 1 time, but with previously unsuspected
relationships in 1T-physics that are not apparent in the ordinary formulation
of physics. Hidden relations among 1T-physics systems, predicted by
2T-physics, provide the observable clues and evidence of the underlying $4+2$
nature of spacetime.

In the present paper we discuss some such relationships in the context of
field theory and provide simple examples of the type of phenomena described
above. These are dualities among 1T field theories in different gravitational
backgrounds (different 1T spacetimes).

The Weyl and general coordinate transformations that relate the field theories
discussed here are familiar transformations and the techniques are buried in
old literature. But these transformations were not previously presented as
duality transformations, nor were they understood to be part of gauge
symmetries that unite the 1T-shadows into a single higher dimensional
structure described by a parent 2T theory. We emphasize that the specific
physics examples discussed explicitly here were not all familiar as being
related by dualities.

We also stress that these simple examples form only a subset of a much larger
set of shadows that obey more complicated duality transformations (not just
Weyl and general coordinate) which were not known to exist until discovered
through 2T-physics.

In this context, the usual Standard Model of Particles and Forces (SM) in 3+1
flat spacetime is regarded as one of the shadows of a parent field theory in
4+2 dimensions. According to our arguments it is dual to a variety of shadows,
some of which are obtained by a series of Weyl and general coordinate
transformations. It may be significant that one of the dual shadows is the SM
in the Robertson-Walker expanding universe.

\section{2T Physics}

While theories with extra spacelike dimensions have been discussed
extensively, theories with more than one timelike dimension have been largely
left aside. M-theory itself as well as its extensions have provided various
signals through supersymmetry stuctures and dualities that extra timelike
dimensions could be relevant for an eventual understanding of fundamental
physics \cite{Stheory}-\cite{sezgin}. However, it is not an easy step to
construct a theory with full fledged extra timelike dimensions due to
interpretational issues and most importantly because of the systematic
presence of ghosts in the quantum theory. Even the first timelike dimension
potentially introduces ghosts in relativistic quantum field or string
theories. Experience over half a century shows that the cure, to remove the
ghosts due to the first timelike dimension, lies in having the right mix of
gauge symmetries to arrive at a unitary and physical theory.

Similarly, Two-Time Physics, in $d$ space and $2$ time dimensions, is a
general framework for a unitary and physical theory, which is achieved
precisely by having the right mix of gauge symmetries. The key element of
2T-physics is the presence of a worldline Sp$\left(  2,R\right)  $ gauge
symmetry which acts in \textit{phase space} $\left(  X^{M},P_{M}\right)  $ and
makes position and momentum indistinguishable at any worldline instant
\cite{2T basics}. \ This Sp$\left(  2,R\right)  $ is an upgrade of worldline
reparametrization to a higher gauge symmetry. It yields nontrivial physical
content only if the target spacetime includes two time dimensions, and plays a
crucial role to remove all unphysical degrees of freedom in a 2T spacetime,
just as worldline reparametrization removes unphysical degrees of freedom in a
1T spacetime.\

In the case of spinning particles the worldline gauge symmetry is extended to
OSp$\left(  n|2\right)  $ \cite{spin2t}\cite{2T field paper} while adding
fermionic spin degrees of freedom $\psi^{M}$ in $d+2$ dimensions beyond phase
space $\left(  X^{M},P_{M}\right)  .$ Similarly, for more complicated systems,
such as supersymmetric particles and others \cite{super2t}-\cite{spinBO}, as
more degrees of freedom with potential ghosts in $d+2$ dimensions are added,
the corresponding worldline gauge symmetry is also larger, to insure the
unitarity of the theory in the 2T-physics formulation. All extensions of the
worldline gauge symmetry must include the key ingredient Sp$\left(
2,R\right)  $, and hence is required to have 2 times.

2T-physics is elevated from the worldline formulation to field theory through
the process of covariant quantization. The spin 0,$\frac{1}{2}$,1 fields,
$\Phi\left(  X\right)  ,\Psi_{\alpha}\left(  X\right)  ,A_{M}\left(  X\right)
,$ are then identified with the first quantized wavefunctions that obey the
gauge symmetry constraints, implying that these fields describe the ghost-free
\textit{gauge invariant} sector of the worldline theory, as long as the
constraints are satisfied as on-shell equations of motion. 2T field theory is
based on an action principle that generates these constraints as equations of
motion, and furthermore extends them with interactions.

The 2T-physics field theory formalism has some features that differ from
1T-field theory formalism, such as a delta function in the volume element
$\delta\left(  X^{2}\right)  d^{d+2}X$ and other properties \cite{2T SM}, as
outlined below. Thanks to these properties, minimizing the 2T field theory
action leads to field equations that reproduce the Sp$\left(  2,R\right)  $ or
other gauge symmetry constraints of the underlying worldline action, thus
insuring the unitarity of the theory.

In this 2T field theory setup, it has been shown that the usual 1T-physics
Standard Model of Particles and Forces in 3+1 dimensions is reproduced as one
of the shadows of a 2T-physics field theory in 4+2 dimensions \cite{2T SM}.
The \textit{emergent 1T Standard Model}, being a 3+1 shadow of the 4+2 theory
with more symmetry, comes with some additional restrictions that are not
present in the usual 1T formulation, but nevertheless agrees with all known
physics. The differences occur only in hitherto unmeasured parts of the
Standard Model, in particular the axion and Higgs sectors, so they are of
phenomenological as well as theoretical significance, and may provide tests at
the LHC or in Cosmology to distinguish 2T-physics from previous approaches.

There are more ways to test 2T-physics at all scales of physics by exploring
the multiple 1T-physics shadows and the predicted relationships among them as
well as their hidden symmetries that give information on the higher
dimensions. Previous work in the context of the worldline formalism displayed
many examples of these shadows \cite{gauge1}-\cite{gauge3}. A graphical
display of some of these examples can be found at \cite{shadows}. In our
recent paper \cite{paper1} most of the known shadows were tabulated and useful
mathematical formulas that describe them were summarized (see tables I,II and
III and related discussion in \cite{paper1}).

This avenue of investigation is still in its infancy. The purpose of our paper
is to develop some techniques and concepts along this path by elucidating the
dualities and hidden symmetries among a subset of these shadows. This subset
is represented by 1T field theories in different gravitational backgrounds
which are all conformally flat. In our recent paper \cite{paper1} the
dualities and hidden symmetries of a 1T scalar field theory in such
backgrounds was discussed. In the present paper we further elaborate on these
properties with fermionic fields that carry spin 1/2, and Yang-Mills gauge
fields that carry spin 1. It is then possible to discuss a subset of the
dualities and hidden symmetries for the Standard Model. We expect that these
dualities, together with the future extension of our results to other types of
shadows, to be potentially useful for non-perturbative analysis of the
Standard Model.

\section{2T field theory}

2T field theory has been fully formulated at the action level for fields of
spins $0,\frac{1}{2},1$ \cite{2T SM}, and to the field equation of motion
level for spin 2 \cite{2T field paper} and beyond \cite{2tbacgrounds}, and has
also been supersymmetrized \cite{susy2tN1}. \ The scalar field was discussed
extensively in our recent paper \cite{paper1}. \ In the current paper, we will
focus on the spin-$\frac{1}{2}$ and spin-1 cases.

\subsection{Spin-$1$ fields}

The 2T action for spin-1 Yang-Mills fields is
\begin{equation}
S\left(  A\right)  =Z\int d^{\left(  d+2\right)  }X~\delta\left(
X^{2}\right)  \left(  -\frac{1}{4}\Phi^{\frac{2\left(  d-4\right)  }{d-2}%
}Tr\left(  F_{MN}F^{MN}\right)  \right)  \label{spin 1 action}%
\end{equation}
where $Z$ is an overall normalization constant that will be determined. \ The
dilaton $\Phi,$ which drops out when $d=4$ in the above expression, is
necessary when $d\neq4$\ for consistency of constraints or 2T gauge symmetries
(see \cite{2T SM}). The action for the dilaton $S\left(  \Phi\right)  $ and
its duality properties have already been discussed in our previous paper
\cite{paper1} that described any scalar, including the dilaton. \ Turning to
the matrix valued Yang-Mills gauge field $A_{M}$ in the adjoint representation
of the gauge group $G,$ the field strength $F_{MN}$ is defined as usual%
\begin{equation}
F_{MN}\equiv\partial_{M}A_{N}-\partial_{N}A_{M}-ig\left[  A_{M},A_{N}\right]
.
\end{equation}
Varying the action with respect to the matrix $A_{N}$ results in the
expression%
\begin{equation}
\delta S\left(  A\right)  =Z\int d^{\left(  d+2\right)  }X~Tr~\left\{  \delta
A_{N}\left[
\begin{array}
[c]{c}%
\delta\left(  X^{2}\right)  D_{M}\left(  \Phi^{\frac{2\left(  d-4\right)
}{d-2}}F^{MN}\right) \\
+2\delta^{\prime}\left(  X^{2}\right)  ~\Phi^{\frac{2\left(  d-4\right)
}{d-2}}X_{M}F^{MN}%
\end{array}
\right]  \right\}  \label{delSA}%
\end{equation}
where $\delta^{\prime}\left(  X^{2}\right)  $ emerges from an integration by
parts. Since the delta function $\delta\left(  X^{2}\right)  $ and its
derivative $\delta^{\prime}\left(  X^{2}\right)  $ are linearly independent
distributions, minimizing the action $\delta S\left(  A\right)  =0$ for
general $\delta A_{N}$ gives two separate equations of motion for $A_{M}$%
\begin{equation}
\left[  X^{N}F_{MN}\right]  _{X^{2}=0}=0,\;\left[  D_{M}\left(  \Phi
^{\frac{2\left(  d-4\right)  }{d-2}}F^{MN}\right)  \right]  _{X^{2}=0}=0.
\label{eom spin 1}%
\end{equation}
The two conditions $X^{2}=0$ and $\left[  X^{N}F_{MN}\right]  _{X^{2}=0}=0$
have been called \textquotedblleft kinematical\textquotedblright\ constraints
\cite{2T SM} that parallel two of the worldline Sp$\left(  2,R\right)  $
constraints $X^{2}=X\cdot P=0$ (applied on states P is a derivative). The
remaining \textquotedblleft dynamical\textquotedblright\ equation of motion
that contains two derivatives parallels the third Sp$\left(  2,R\right)  $
worldline constraint $P^{2}=0$. The field theoretic version of these
Sp$\left(  2,R\right)  $ constraints\footnote{Taking into consideration the
spin degrees of freedom carried by the vector field $A_{M}\left(  X\right)  $,
the full set of constraints is actually OSp$\left(  2|2\right)  ,$ where
OSp$\left(  2|2\right)  $ is the gauge symmetry of the worldline theory for a
spin 1 particle \cite{spin2t}\cite{2T field paper}.} evidently include field
interactions that are consistent with the familiar Yang-Mills gauge symmetry.

The delta function $\delta\left(  X^{2}\right)  $ that appears in the action
invites an expansion of every field in powers of $X^{2}.$ For the gauge field
one can write
\begin{equation}
A_{M}=A_{M}^{0}+X^{2}\tilde{A}_{M} \label{expand}%
\end{equation}
where we define $A_{M}^{0}\equiv\left[  A_{M}\right]  _{X^{2}=0}$ while
$X^{2}\tilde{A}_{M}=A_{M}-A_{M}^{0}$ is the remainder that includes all higher
powers of $X^{2}$. As shown in \cite{2T SM}, the action $S\left(  A\right)  $
has also a \textquotedblleft2T-gauge symmetry\textquotedblright\ under the
variation%
\begin{equation}
\delta_{\Lambda}A_{N}=\Phi^{-\frac{2\left(  d-4\right)  }{d-2}}X^{2}%
\Lambda_{N}\left(  X\right)  \label{gaugeSymA}%
\end{equation}
which can be verified (with some restrictions\footnote{In \cite{2T SM} the 2T
gauge symmetry was discussed under the assumption that the \textit{remainder}
$\tilde{A}_{M}\left(  X\right)  $ in Eq.(\ref{expand}) \`{a} priori satisfied
a homogeneity condition $\left(  X\cdot D+3\right)  \tilde{A}_{M}=0$ (but
unrestricted $A_{M}^{0}$). This condition on $\tilde{A}_{M}\left(  X\right)  $
was a partial gauge choice for a larger gauge symmetry, and therefore the
gauge parameter $\Lambda_{M}\left(  X\right)  $ was also restricted by a
corresponding homogeneity condition $\left(  X\cdot D+3\right)  \Lambda_{M}%
=0$. A homogeneous $\tilde{A}_{M}\left(  X\right)  $ made it easier to derive
the two separate equations in (\ref{eom spin 1}) as the unique outcome of
minimizing the action. This assumption for $\tilde{A}_{M}\left(  X\right)  $
can be dropped at the expense of a more elaborate discussion of the larger 2T
gauge symmetry, as will be further elucidated in a separate paper. With this,
one arrives again at the same on-shell equations of motion (\ref{eom spin 1}).
Either way, the conclusions of the present paper remain unchanged.} on the
local gauge parameter $\Lambda_{M}\left(  X\right)  $) by inserting
$\delta_{\Lambda}A_{N}$ into Eq.(\ref{delSA}) instead of the general $\delta
A_{N}$. This gauge symmetry can be used to thin out the degrees of freedom in
$A_{M}\left(  X\right)  .$ In \cite{2T SM} it was argued that there is just
enough \textquotedblleft2T-gauge symmetry\textquotedblright\ to remove the
remainder $\tilde{A}_{M}\left(  X\right)  $ if so desired, thus showing that
the gauge fixed fields become independent of $X^{2}.$ This amounts to
eliminating one spacetime coordinate among the $X^{M}.$

The strategy to descend to 1T-physics from 2T-physics is then to make gauge
choices and solve the two kinematic constraints $X^{2}=0$, $\left[
X^{N}F_{MN}\right]  _{X^{2}=0}=0.$ Upon inserting the solution into the
dynamical field equation or into the original action, one realizes that the
remaining dynamics is in one less space and one less time dimensions precisely
as in 1T-physics field theory, but in a variety of spacetimes. This is then
how we obtain many 1T shadows of the 2T field theory.

The interesting phenomena are that there are many Yang-Mills 1T shadows in
different emerging 1T spacetimes that materialize from different solutions of
the kinematic equations $X^{2}=0$, $\left[  X^{N}F_{MN}\right]  _{X^{2}=0}=0$,
and that the emergent 1T field theories may come with some symmetry
restrictions that are not anticipated with only 1T field theory methods. For
example, in the case of the Standard Model \cite{2T SM} the latter
restrictions lead to new ways of solving the strong CP violation problem
without an axion, as well as to new concepts on the generation of mass.

As already mentioned, among the many possible solutions, in the next section
we will concentrate on an easier subset of solutions that correspond to
conformally flat spacetimes and then explore the dualities among the resulting
field theories.

\subsection{Spin-$\frac{1}{2}$ fields}

The 2T free field action for spinor fields is given by \cite{2T SM}%
\begin{equation}
S\left(  \Psi\right)  =\frac{i}{2}Z\int\left(  d^{d+2}X\right)  \delta\left(
X^{2}\right)  \left(  \bar{\Psi}\not X  \bar{\not \partial }\Psi+\bar{\Psi
}\overleftarrow{\not \partial }\bar{\not X  }\Psi\right)
\label{spinor action}%
\end{equation}
where $Z$ is the same normalization constant as in (\ref{spin 1 action})$,$
and $\not X  \equiv\Gamma^{M}X_{M}$ and $\bar{\not \partial }\equiv\bar
{\Gamma}^{M}\partial_{M},$ using the SO$\left(  d,2\right)  $ gamma matrix
conventions for $\Gamma^{M},\bar{\Gamma}^{M}$ in the appendix of
\cite{susy2tN1}. Varying the action gives%
\begin{equation}
\delta S\left(  \Psi\right)  =iZ\int\left(  d^{d+2}X\right)  \delta\left(
X^{2}\right)  \delta\bar{\Psi}\left[  \not X  \bar{\not \partial }\Psi-\left(
X\cdot\partial+\frac{d}{2}\right)  \Psi\right]  +h.c. \label{delSpsi}%
\end{equation}
where the second term emerges from integration by parts and using $X^{2}%
\delta^{\prime}\left(  X^{2}\right)  =-\delta\left(  X^{2}\right)  .$ As was
shown in \cite{2T SM}, the two terms in the bracket actually need to vanish
separately when we require $\delta S\left(  \Psi\right)  =0$ for general
$\delta\bar{\Psi}.$ So the equations of motion are\footnote{These equations of
motion amount to OSp$\left(  1|2\right)  $ constraints \cite{2T SM}, where
OSp$\left(  1|2\right)  $ is the gauge symmetry\cite{2T field paper} of the
underlying worldline theory \cite{spin2t} (see Appendix
(\ref{spin half worldine})). Imposing OSp$\left(  1|2\right)  $ constraints is
the requirement that the physical configurations of the field $\Psi\left(
X\right)  $ be gauge invariant under the OSp$\left(  1|2\right)  $ gauge
symmetry.}%
\begin{equation}
\left[  \left(  X\cdot\partial+\frac{d}{2}\right)  \Psi\right]  _{X^{2}%
=0}=0,\;\;\left[  \not X  \bar{\not \partial }\Psi\right]  _{X^{2}=0}=0.
\label{eom spin half}%
\end{equation}

It should be noted that the action $S\left(  \Psi\right)  $ is invariant under
the following \textquotedblleft2T gauge transformation\textquotedblright\
\begin{equation}
\delta_{\zeta}\Psi=X^{2}\zeta_{1}+\not X  \zeta_{2},\;\delta_{\zeta}\bar{\Psi
}=X^{2}\bar{\zeta}_{1}+\bar{\zeta}_{2}\bar{\not X  }.
\label{spin half gauge sym}%
\end{equation}
This is verified (see \cite{2T SM}) by inserting $\delta_{\zeta}\bar{\Psi}$ in
(\ref{delSpsi}) instead of the general $\delta\bar{\Psi}.$ The role of the
gauge spinors $\zeta_{1},\zeta_{2}$ are as follows. Due to the delta function
we are invited to expand the field in powers of $X^{2},$ thus $\Psi=\Psi
^{0}+X^{2}\tilde{\Psi},$ where we define $\Psi^{0}\equiv\left[  \Psi\right]
_{X^{2}=0}$ while $X^{2}\tilde{\Psi}=\Psi-\Psi^{0}$ is the remainder that
includes all higher powers of $X^{2}$. In \cite{2T SM} it was shown that the
gauge parameter $\zeta_{1}$ that appears in Eq.(\ref{spin half gauge sym}) can
be used to remove the remainder $\tilde{\Psi}$ if so desired. The remaining
$\Psi_{\alpha}^{0}\left(  X\right)  $ is then independent of $X^{2},$ however
compared to 2 less dimensions it has double the number of spinor components.
With the gauge symmetry $\zeta_{2}$ one can show \cite{2T SM} that half of the
degrees of freedom in $\Psi_{\alpha}^{0}\left(  X\right)  $ are gauge degrees
of freedom while the remaining half are physical. In this role, the $\zeta
_{2}$ transformation is similar to kappa-type local supersymmetry, and it can
be used to eliminate half of the spinor components, if so desired.

Interactions of fermions with the gauge fields are obtained by simply
replacing all derivatives by the covariant derivative $\partial_{M}\rightarrow
D_{M}=\partial_{M}-igA_{M}.$ The Yukawa interaction with a scalar $H\left(
X\right)  $ takes the form \cite{2T SM} $H\left(  \bar{\Psi}{\not X  }%
\Psi\right)  \Phi^{\frac{2\left(  d-4\right)  }{\left(  d-2\right)  }},$ where
$\Phi$ is the dilaton that does not appear if $d+2=6$. The fermionic gauge
symmetry of Eq.(\ref{spin half gauge sym}) remains as a valid symmetry in the
presence of these interactions, and it will be used to obtain the proper
spin-$\frac{1}{2}$ degrees of freedom in the lower dimensional actions.

The strategy to descend to 1T-physics from 2T-physics for fermions is then to
make gauge choices by using $\zeta_{1},\zeta_{2}$ and solve the two kinematic
constraints $X^{2}=0$ and ($(X\cdot D+\frac{d}{2})\Psi)_{X^{2}=0}=0.$ Upon
inserting the solution into the original action (including interactions) it is
seen that the remaining dynamics has precisely the familiar form of 1T field theory.

As in the case of gauge fields above, various 1T spacetimes materialize from
different solutions of the kinematic equations. These emerging 1T field
theories in $\left(  d-1\right)  +1$ dimensions, that include scalars,
fermions and Yang-Mills bosons, are then dual to each other. This duality will
be illustrated below for a subset of the solutions.

\section{Emergent $(d-1)+1$ field theory}

The strategy described in the previous section to reduce 2T field theory to 1T
field theory will be implemented in this section by solving the kinematic
equations
\begin{equation}
X^{2}=0,\;\;\left[  (X\cdot D+\frac{d}{2})\Psi\right]  _{X^{2}=0}%
=0,\;\;\left[  X^{N}F_{MN}\right]  _{X^{2}=0}=0. \label{kinematics}%
\end{equation}
The result, which will involve fields in 2 less spacetime variables, will be
inserted in the original action to yield the \textquotedblleft
shadows\textquotedblright\ in the form of 1T field theories. To solve these
equations we follow the footsteps for solving the corresponding constraints
$X^{2}=X\cdot P=0$ in the underlying worldline theory. This involved making
some gauge choices for phase space $\left(  X^{M}\left(  \tau\right)
,P_{M}\left(  \tau\right)  \right)  $ by using the worldline local Sp$\left(
2,R\right)  $ gauge symmetry. In this way the 1T systems listed in Table-1
emerge as \textquotedblleft shadows\textquotedblright\ from the 2T theory in
flat $d+2$ dimensions. In this table the cases marked as $\left(  cf\right)  $
correspond to conformally flat curved spaces, on which we concentrate in this
paper.
\[%
\begin{tabular}
[c]{|l|}\hline
The massless relativistic particle in $d$ flat Minkowski space.$(cf)$\\
The massive relativistic particle in $d$ flat Minkowski space.\\
The nonrelativistic free massive particle in $d-1$ space dimensions.\\
The nonrelativistic hydrogen atom (i.e. $1/r$ potential) in $d-1$ space
dimensions.\\
The harmonic oscillator in $d-2$ space dimensions, with its mass
$\Leftrightarrow$ an extra dimension.\\
The particle on AdS$_{d}$, or on dS$_{d}.(cf)$\\
The particle on AdS$_{d-k}\times$S$^{k}$ for $k=1,2,\cdots,d-1.$ $(cf)$\\
The particle on the Robertson-Walker spacetime (open or closed universes).
$(cf)$\\
The particle on any maximally symmetric space of positive or negative
curvature. $(cf)$\\
The particle on any of the above spaces modified by any conformal factor.\\
A related family of other particle systems, including some singular
backgrounds.$(cf)$\\\hline
$\text{Table1 - A sample of 1T physics \textquotedblleft
shadows\textquotedblright\ that emerge from the flat }\left(  d+2\right)
\text{ 2T theory.}$\\\hline
\end{tabular}
\ \
\]
The details of the worldline gauge choices for $\left(  X^{M},P_{M}\right)  $
was summarized in tables I,II,III in \cite{paper1}. Those tables provide
details for a variety of embeddings of $\left(  d-1\right)  +1$ dimensions
into $d+2$ dimensions, with distinct forms of \textquotedblleft
time\textquotedblright\ and \textquotedblleft Hamiltonian\textquotedblright%
\ as interpreted in the lower dimension (i.e. the 1T shadows).

In 2T field theory, we cannot choose a gauge\footnote{This is because
Sp$\left(  2,R\right)  $ is not a gauge symmetry of the \textit{field theory}
action $S\left(  \Phi,A,\Psi\right)  $, but rather the action generates
on-shell equations of motion that reproduce the Sp$\left(  2,R\right)  $
constraints of the \textit{worldline theory}, as explained in the previous
section. These fields which satisfy the Sp$\left(  2,R\right)  $ constraints
are then the Sp$\left(  2,R\right)  $ gauge invariant physical
configurations.} for $X^{M}$ like we do for the worldline theory $X^{M}\left(
\tau\right)  $. Instead, we \textit{parameterize} $X^{M}$ as in e.g.
Eq.(\ref{parametrization}), which is a form that is parallel to a subset of
gauge choices of the worldline theory (compare to Appendix
\ref{spin half worldine}). \ We start by choosing an embedding of the 1T
spacetime $x^{\mu}$ into the 2T spacetime $X^{M}.$ To do so, it is useful to
distinguish one space and one time dimensions $X^{0^{\prime}},X^{1^{\prime}},$
to define a lightcone-type basis, $\ M=\left(  +^{\prime},-^{\prime},m\right)
$, with $X^{\pm^{\prime}}\equiv\frac{1}{\sqrt{2}}\left(  X^{0^{\prime}}\pm
X^{1^{\prime}}\right)  $, so that the flat metric $\eta_{MN}$ in $d+2$
dimensions takes the form%
\begin{equation}
ds^{2}=dX^{M}dX^{N}\eta_{MN}=-2dX^{+^{\prime}}dX^{-^{\prime}}+dX^{m}dX^{n}%
\eta_{mn} \label{flat}%
\end{equation}
where $\eta_{mn}$ is the flat Minkowski metric in $d$ dimensions including 1
time dimension. Next we choose the embedding by the following
\textit{parametrization} of $X^{M}$ in terms of the 1T spacetime $x^{\mu}$ and
two other dimensions $\kappa,\lambda$%
\begin{equation}
X^{+^{\prime}}=\kappa e^{\sigma\left(  x\right)  },\;X^{-^{\prime}}%
=\lambda\kappa e^{\sigma\left(  x\right)  },\;X^{m}=\kappa e^{\sigma\left(
x\right)  }q^{m}\left(  x\right)  , \label{parametrization}%
\end{equation}
where the functions $\sigma\left(  x\right)  $ and $q^{m}\left(  x\right)  $
remain unspecified. Solving for $\kappa$, $\lambda$ and $q^{m}\left(
x\right)  $, in terms of $X^{\pm^{\prime}},X^{m}$ we get the inverse
parametrization%
\begin{equation}
\kappa=e^{-\sigma\left(  x\right)  }X^{+^{\prime}},\;\;\lambda=\frac
{X^{-^{\prime}}}{X^{+^{\prime}}},\;\;q^{m}\left(  x\right)  =\frac{X^{m}%
}{X^{+^{\prime}}}. \label{q}%
\end{equation}
From $q^{m}\left(  x\right)  =\frac{X^{m}}{X^{+\prime}}$ we solve in principle
for $x^{\mu}=f^{\mu}\left(  \frac{X^{m}}{X^{+\prime}}\right)  ,$ where
$f^{\mu}\left(  q^{m}\right)  $ is the inverse map of $q^{m}\left(  x^{\mu
}\right)  .$ This inverse map is inserted in $\sigma\left(  x\right)  =$
$\sigma\left(  f^{\mu}\left(  \frac{X^{m}}{X^{+\prime}}\right)  \right)  $ in
Eq.(\ref{q}) to complete the full solution of $\kappa=X^{+^{\prime}}%
\exp\left(  -\sigma\left(  f^{\mu}\left(  \frac{X^{m}}{X^{+\prime}}\right)
\right)  \right)  $ in terms of $X^{\pm^{\prime}},X^{m}.$

Such parametrizations of $X^{M}$, combined with gauge choices for Yang-Mills
gauge symmetry and 2T gauge symmetries (\ref{gaugeSymA}%
,\ref{spin half gauge sym}), lead to the solutions of Eqs.(\ref{kinematics})
as will be shown below.

The physics of the emerging 1T shadows as field theories is anticipated from
the corresponding shadows in the classical worldline theory. The improvements
in field theory include (i) an automatic resolution of ordering ambiguities of
nonlinear terms in the quantization of the worldline theory (see Appendix
\ref{spin half worldine}), (ii) the inclusion of interactions and (iii)
dualities among interacting field theories which may be used as a new tool for
investigating 1T field theory.

To implement the 2T$\rightarrow$1T reduction for spin-$\frac{1}{2}$ and
spin-$1$ fields we solve Eqs.(\ref{kinematics}). We follow the methods of our
previous investigation of scalar fields \cite{paper1} which focused on
conformally flat 1T-spacetimes that emerged through the 2T$\rightarrow$1T
embeddings described by Eq.(\ref{parametrization}). The conformally flat
backgrounds, which is only a subset of the \textquotedblleft
shadows\textquotedblright\ listed in Table-1, are those markes as $\left(
cf\right)  $ including the flat massless Minkowski spacetime, AdS$_{d-k}%
\times$S$^{k},$ AdS$_{d}$, dS$_{d}$, Robertson-Walker, maximally symmetric
spaces, and some singular spaces. The other interesting cases listed in
Table-1, such as the massive particle(s), hydrogen atom, and harmonic
oscillator are not conformally flat backgrounds. To describe those other
non-conformally flat \textquotedblleft shadows\textquotedblright, which are
also solutions of Eqs.(\ref{kinematics}),\ a parametrization of the embedding
of the 1T spacetime into the 2T spacetime that is rather different than
Eq.(\ref{parametrization}) is required\footnote{The special case treated in
this paper in Eq.(\ref{parametrization}) embeds $d$ dimesional space $x^{\mu}$
space into $d+2$ dimensional space $X^{M}$, from which we can figure out the
embedding of momentum (derivatives applied on fields as in Eqs.(\ref{chain1}%
-\ref{chain3})). The more general case embeds not only space $x^{\mu}$, but
all of all of phase space $\left(  x^{\mu},p_{\mu}\right)  $ in $d$ dimensions
into phase space $\left(  X^{M},P_{M}\right)  $ in $d+2$ dimensions.
Consequently, the emergent spacetimes are not only conformally flat, but much
more interesting. Some such examples include the massive particle(s), the
hydrogen atom and harmonic oscillator listed in Table-1. In these later cases
the parametrization of $X^{M}$ involves momenta in addition to positions (see
tables I,II,III in \cite{paper1}). While this is straightforward to implement
in the worldline formalism, it is more challenging in the context of field
theory, since momenta are replaced by derivatives. For this reason, the field
theoretic investigation of this more complicated type of \textquotedblleft
shadow\textquotedblright\ is left to future work.\label{others}}.

The fields $\Phi\left(  X\right)  =\Phi\left(  \kappa,\lambda,x^{\mu}\right)
,~\Psi\left(  X\right)  =\Psi\left(  \kappa,\lambda,x^{\mu}\right)  $ and
$A_{M}\left(  X\right)  =A_{M}\left(  \kappa,\lambda,x^{\mu}\right)  $ are now
considered functions of $\kappa,\lambda,x^{\mu}.$ The $\left(  d-1\right)  +1$
spacetime $x^{\mu}$ has been embedded in $d+2$ dimensions in different forms
that vary as the functions $q^{m}\left(  x\right)  $ and $\sigma\left(
x\right)  $ change.

To obtain the kinetic energy terms for the fields $\Phi\left(  X\right)
,A_{M}\left(  X\right)  ,\Psi_{\alpha}\left(  X\right)  $ we use the chain
rule to compute the partial derivatives $\frac{\partial}{\partial X^{M}}$ in
terms of $\frac{\partial}{\partial\kappa},\frac{\partial}{\partial\lambda
},\frac{\partial}{\partial x^{\mu}},$ consistent with the parametrization
(\ref{parametrization}). The result is%
\begin{align}
\frac{\partial}{\partial X^{-^{\prime}}}  &  =\frac{1}{\kappa}e^{-\sigma}%
\frac{\partial}{\partial\lambda}\label{chain1}\\
\frac{\partial}{\partial X^{m}}  &  =\frac{1}{\kappa}\left(  -e_{m}^{\mu
}\partial_{\mu}\sigma~\kappa\frac{\partial}{\partial\kappa}+e_{m}^{\mu
}\partial_{\mu}\right) \label{chain2}\\
\frac{\partial}{\partial X^{+^{\prime}}}  &  =\frac{1}{\kappa}\left(  \left[
e^{-\sigma}+q^{m}e_{m}^{\mu}\partial_{\mu}\sigma\right]  ~\kappa\frac
{\partial}{\partial\kappa}-e^{-\sigma}\lambda\frac{\partial}{\partial\lambda
}-q^{m}e_{m}^{\mu}\partial_{\mu}\right)  \label{chain3}%
\end{align}
Here $e_{m}^{\mu}\left(  x\right)  $ is the inverse of the vielbein. The
vielbein itself in the reduced spacetime is defined as%
\begin{equation}
e_{\mu}^{m}\left(  x\right)  =e^{\sigma\left(  x\right)  }\frac{\partial
q^{m}\left(  x\right)  }{\partial x^{\mu}}, \label{vilebein}%
\end{equation}
Then the inverse $e_{m}^{\mu}\left(  x\right)  $ can also be written as
$e_{m}^{\mu}\left(  x\right)  =e^{-\sigma\left(  x\right)  }\frac{\partial
x^{\mu}}{\partial q^{m}}=e^{-\sigma\left(  x\right)  }\frac{\partial f^{\mu
}\left(  q\right)  }{\partial q^{m}}\left(  x\right)  ,$ where $x^{\mu}%
=f^{\mu}\left(  q\right)  $ is the inverse map discussed following
Eq.(\ref{q}). This is verified by using the chain rule $e_{\nu}^{~m}\left(
x\right)  e_{m}^{\mu}\left(  x\right)  =e^{\sigma\left(  x\right)  }%
\frac{\partial q^{m}}{\partial x^{\nu}}e^{-\sigma\left(  x\right)  }%
\frac{\partial x^{\mu}}{\partial q^{m}}=\frac{\partial x^{\mu}}{\partial
x^{\nu}}=\delta_{\nu}^{~\mu}.$ We note in particular that the dimension
operator $X\cdot\partial$ that we will need to solve the kinematic equations
(\ref{kinematics}) takes a simple form $\kappa\frac{\partial}{\partial\kappa
}$
\begin{equation}
X\cdot\partial=-X^{+^{\prime}}\frac{\partial}{\partial X^{-^{\prime}}%
}-X^{-^{\prime}}\frac{\partial}{\partial X^{+^{\prime}}}+X^{m}\frac{\partial
}{\partial X^{m}}=\kappa\frac{\partial}{\partial\kappa}. \label{dimOp}%
\end{equation}

With this parametrization we see that the volume element takes the form%
\begin{align}
X^{2}  &  =-2\kappa^{2}e^{2\sigma}\left(  \lambda-\frac{1}{2}q^{2}\left(
x\right)  \right)  ,\\
\left(  d^{d+2}X\right)  ~\delta\left(  X^{2}\right)   &  =\frac{1}{2}%
\kappa^{d-1}\det\left(  e_{\mu}^{m}\left(  x\right)  \right)  ~d\kappa
d\lambda d^{d}x~\delta\left(  \lambda-\frac{1}{2}q^{2}\left(  x\right)
\right)  . \label{volume}%
\end{align}
where we have taken into account the Jacobian for the change of variables
\begin{equation}
J\left(  \frac{X^{+^{\prime}},X^{-^{\prime}},X^{m}}{\kappa,\lambda,x^{\mu}%
}\right)  =\kappa^{d+1}e^{\left(  d+2\right)  \sigma}\det\left(  \partial
_{\mu}q^{m}\right)  =\kappa^{d+1}e^{2\sigma}\det\left(  e_{\mu}^{m}\left(
x\right)  \right)  .
\end{equation}
It is also worth noticing that, after taking into account the delta function
that imposes $\lambda=\frac{1}{2}q^{2}\left(  x\right)  ,$ the metric in $d+2$
dimensions in Eq.(\ref{flat}) collapses to the curved metric $g_{\mu\nu
}\left(  x\right)  $ in $\left(  d-1\right)  +1$ dimensions%
\begin{equation}
ds^{2}=\left(  dX\cdot dX\right)  _{\lambda=q^{2}/2}=\kappa^{2}g_{\mu\nu
}\left(  x\right)  dx^{\mu}dx^{\nu}. \label{flat2}%
\end{equation}
Here $g_{\mu\nu}\left(  x\right)  $ is conformally flat since it has the form%
\begin{equation}
g_{\mu\nu}=e_{\mu}^{m}e_{\nu}^{n}\eta_{mn}=e^{2\sigma\left(  x\right)  }%
\frac{\partial q^{m}\left(  x\right)  }{\partial x^{\mu}}\frac{\partial
q^{n}\left(  x\right)  }{\partial x^{\nu}}\eta_{mn}. \label{metric1}%
\end{equation}

Specific forms of $\sigma\left(  x^{\mu}\right)  ,q^{m}\left(  x^{\mu}\right)
$ that produce all of the conformally flat examples included in Table-1 were
given explicitly in \cite{paper1}. As an illustration, we give here the
Robertson-Walker case, where $a\left(  t\right)  $ is any function that
represents the expanding size of an open universe%
\begin{equation}
\left(  ds^{2}\right)  _{\lambda=q^{2}/2}=\frac{\kappa^{2}}{R_{0}^{2}}\left[
-dt^{2}+\frac{a^{2}\left(  t\right)  }{R_{0}^{2}}\left(  \frac{R_{0}^{2}%
}{R_{0}^{2}+r^{2}}dr^{2}+r^{2}d\Omega^{2}\right)  \right]
\end{equation}
This is an example of Eq.(\ref{flat2}) that is obtained by inserting the
following explicit forms of $\sigma\left(  x\right)  $ and $q^{m}\left(
x\right)  $
\begin{align}
&  \text{Robertson-Walker expanding open universe }\left(  r>0\right)
\nonumber\\
e^{\sigma\left(  x\right)  }  &  \equiv\frac{a\left(  t\right)  }{R_{0}}%
\exp\left(  \pm\int^{t}\frac{dt^{\prime}}{a(t^{\prime})}\right)  ,\;\vec
{q}\left(  x\right)  \equiv\pm\frac{\vec{r}}{R_{0}}\exp\left(  \mp\int
^{t}\frac{dt^{\prime}}{a(t^{\prime})}\right)  ,\\
q^{0}\left(  x\right)   &  \equiv\mp\sqrt{1+\frac{r^{2}}{R_{{\small 0}}^{2}}%
}\exp\left(  \mp\int^{t}\frac{dt^{\prime}}{a(t^{\prime})}\right) \\
e_{\mu}^{~m}  &  =e^{\sigma\left(  x\right)  }\frac{\partial q^{m}}{\partial
x^{\mu}}=\frac{1}{R_{0}}\left(
\begin{array}
[c]{cc}%
\sqrt{1+\frac{r^{2}}{R_{{\small 0}}^{2}}} & ~-\frac{r^{i}}{R_{0}}\\
\mp\frac{a\left(  t\right)  }{R_{0}}\frac{r_{i}}{\sqrt{R_{{\small 0}}%
^{2}+r^{2}}} & \;\;\;\pm\frac{a\left(  t\right)  }{R_{0}}\delta_{i}^{~j}%
\end{array}
\right)
\end{align}
This parametrization of $X^{M}\left(  \kappa,\lambda,x^{\mu}\right)  $ given
above for the Robertson Walker spacetime is slightly different than the one
given in \cite{paper1}, but is related to it by a simple redefinition of coordinates.

The discussion above is common to fields of any spin. \ The 2T$\rightarrow$1T
reduction of the spin 0 case was discussed in \cite{paper1}, so we now focus
on the spin-$\frac{1}{2}$ and spin-1 cases.

\subsection{Spin-1 field}

The kinematic equations (\ref{kinematics}) were first solved by Dirac
\cite{Dirac} (see also related work in \cite{kastrup}-\cite{vasiliev2}) who
did not have an action principle but only suggested equations of motion that
were arrived at by a different set of arguments and the motiovation being an
explanation of conformal symmetry SO$\left(  d,2\right)  $ in flat Minkowski
space in $d$ dimensions. His solution yielded only one of the possible
shadows, namely the one in flat Minkowski space. The existence of the all the
other shadows in a variety of spacetimes, and the existence of moduli such as
curvature, mass, interaction parameters, were discovered via methods of
2T-physics. In what follows we adapt the 2T-physics methods to discuss a
subset of the shadows.

Taking advantage of the Yang-Mills gauge symmetry of the full action $S\left(
\Phi,A,\Psi\right)  ,$ we first choose the Yang-Mills axial gauge
\begin{equation}
X\cdot A=0 \label{axialGauge}%
\end{equation}
so that the non-linear kinematic equation for the Yang-Mills field in
(\ref{kinematics}) simplifies to a linear equation independent of interactions
$0=X^{N}F_{MN}=-\left(  X\cdot\partial+1\right)  A_{M}\left(  X\right)  .$
Using the parametrization for $X^{M}$ in (\ref{parametrization}) that yields
the dimension operator as in (\ref{dimOp}), the kinematic constraint takes the
even simpler form $\ $%
\begin{equation}
\left(  \kappa\frac{\partial}{\partial\kappa}+1\right)  A_{M}\left(
\kappa,\lambda,x\right)  =0.
\end{equation}
This determines uniquely the $\kappa$ dependence of the field as%
\begin{equation}
A_{M}\left(  \kappa,\lambda,x\right)  =\frac{1}{\kappa}\hat{A}_{M}\left(
\lambda,x\right)  .
\end{equation}
In the axial gauge (\ref{axialGauge}) there is still leftover Yang-Mills gauge
symmetry with parameter $\Lambda\left(  \lambda,x\right)  $ that is
independent of $\kappa.$ Using this, we can fix the Yang-Mills gauge further,
by taking a lightcone-type gauge%
\begin{equation}
\hat{A}_{-^{\prime}}\left(  \lambda,x\right)  =0.
\end{equation}
Inserting this in the axial gauge condition $0=X\cdot\hat{A}=X^{-^{\prime}%
}\hat{A}_{-^{\prime}}+X^{+^{\prime}}\hat{A}_{+^{\prime}}+X^{m}A_{m}$ , we also
obtain $\hat{A}_{+^{\prime}}=-\frac{X^{m}}{X^{+^{\prime}}}\hat{A}_{m}$, which
may be written as%
\begin{equation}
\hat{A}_{+^{\prime}}\left(  \lambda,x\right)  =-q^{m}\left(  x\right)  \hat
{A}_{m}\left(  \lambda,x\right)  .
\end{equation}

Given the delta function $\delta\left(  \lambda-\frac{1}{2}q^{2}\left(
x\right)  \right)  $ in the volume element (\ref{volume}) we are invited to
expand the field in powers of $\lambda-\frac{1}{2}q^{2}\left(  x\right)  .$
Now we take advantage of the 2T gauge symmetry of Eq.(\ref{gaugeSymA}) which
allows us to gauge fix the part of the gauge field $A_{M}=A_{M}^{0}%
+X^{2}\tilde{A}$ proportional to $X^{2}.$ This means that in the expansion
$\hat{A}_{m}\left(  \lambda,x^{\mu}\right)  =\bar{A}_{m}\left(  x^{\mu
}\right)  +\left(  \lambda-\frac{q^{2}}{2}\right)  \tilde{A}_{m}\left(
\lambda,x^{\mu}\right)  $ we can choose the gauge $\tilde{A}_{m}\left(
\lambda,x^{\mu}\right)  =0$. \ The remaining gauge field $\bar{A}_{m}\left(
x^{\mu}\right)  $ is now independent of both $\lambda$ and $\kappa.$ Without
loss of generality we can write it in the form $\bar{A}_{m}\left(  x^{\mu
}\right)  =e_{m}^{\mu}\left(  x\right)  A_{\mu}\left(  x\right)  $ where
$e_{m}^{\mu}$ is the inverse vielbein discussed in the previous section.

To summarize, we have shown that, by taking advantage of both the Yang-Mills
and 2T gauge symmetries, the general Yang-Mills field $A_{M}\left(  X\right)
$ can be gauge fixed to the following form%
\begin{equation}
A_{-^{\prime}}=0,\;\;A_{+^{\prime}}=-\frac{1}{\kappa}q^{m}\left(  x\right)
e_{m}^{\mu}\left(  x\right)  A_{\mu}\left(  x\right)  ,\;\;A_{m}=\frac
{1}{\kappa}e_{m}^{\mu}\left(  x\right)  A_{\mu}\left(  x\right)
\label{gaugedA}%
\end{equation}
where only $A_{\mu}\left(  x\right)  $ is the independent component. We now
compute the field strength $F_{MN}\left(  X\right)  $ by using the chain rule
given in Eqs.(\ref{chain1}-\ref{chain3}). After some algebra we
find\footnote{The details for similar steps in flat space (i.e. when $e_{\mu
}^{m}\left(  x\right)  =\delta_{\mu}^{m}$) are given in \cite{2T SM}.}%
\begin{equation}
F_{+^{\prime}-^{\prime}}=0,\;F_{-^{\prime}m}=0,\;F_{+^{\prime}m}=\frac
{1}{\kappa^{2}}q^{n}\left(  x\right)  F_{mn},\;F_{mn}=\frac{1}{\kappa^{2}%
}e_{m}^{\mu}e_{n}^{\nu}F_{\mu\nu}\left(  x\right)
\end{equation}
with%
\begin{equation}
F_{\mu\nu}\left(  x\right)  =\partial_{\mu}A_{\nu}-\partial_{\nu}A_{\mu
}-ig\left[  A_{\mu},A_{\nu}\right]  .
\end{equation}
The quantity $F_{MN}F^{MN}$ in flat $d+2$ dimensions is then reduced to its
shadow in curved $d$ dimensions as follows
\begin{align}
F_{MN}F^{MN}  &  =F_{mn}F^{mn}=F_{mn}F_{kl}\eta^{mk}\eta^{nl}=\frac{1}%
{\kappa^{4}}\left(  e_{m}^{\mu}e_{n}^{\nu}F_{\mu\nu}\right)  \left(
e_{k}^{\kappa}e_{l}^{\lambda}F_{\kappa\lambda}\right)  \eta^{mk}\eta
^{nl}\nonumber\\
&  =\frac{1}{\kappa^{4}}g^{\mu\kappa}g^{\nu\lambda}F_{\mu\nu}F_{\kappa\lambda
}\equiv\frac{1}{\kappa^{4}}F_{\mu\nu}F^{\mu\nu}%
\end{align}
Recalling also the reduced form of the scalar field from \cite{paper1}
\begin{equation}
\Phi\left(  \kappa,\lambda,x^{\mu}\right)  =\kappa^{-\frac{d-2}{2}}\phi\left(
x^{\mu}\right)  ,
\end{equation}
we can rewrite the action in terms of the lower dimensional shadow fields in
curved space (the subscript \textquotedblleft red\textquotedblright\ indicates
that the solution of the kinematic equations are inserted to obtain the
reduced action)%
\begin{align}
S\left(  A,\Phi\right)  _{red}  &  =Z\int d^{\left(  d+2\right)  }%
X\delta\left(  X^{2}\right)  \left(  -\frac{1}{4}\Phi^{\frac{2\left(
d-4\right)  }{d-2}}Tr\left(  F_{MN}F^{MN}\right)  \right)  _{red}%
\label{redAction1}\\
&  =Z\int d\kappa d\lambda d^{d}x\frac{1}{2}\kappa^{d-1}\det e_{\mu}%
^{m}~\delta\left(  \lambda-\frac{q^{2}\left(  x\right)  }{2}\right)  -\frac
{1}{4}\kappa^{4-d}\phi^{\frac{2\left(  d-4\right)  }{d-2}}Tr\left(  \frac
{1}{\kappa^{4}}F_{\mu\nu}F^{\mu\nu}\right) \nonumber\\
&  =\int d^{d}x\sqrt{-g}\left(  -\frac{1}{4}\phi^{\frac{2\left(  d-4\right)
}{d-2}}Tr\left(  F_{\mu\nu}F^{\mu\nu}\right)  \right)  . \label{redAction3}%
\end{align}
In the last step we integrated $\lambda,\kappa$ and absorbed an infinite
constant by normalizing $Z$ as $Z\int\frac{d\kappa}{2\kappa}=1$, thus arriving
at an action expressed in terms of only the $d$ dimensional shadow spacetime
$x^{\mu}$.

It must be emphasized that the reduction of the scalar field\footnote{There
can of course be a variety of scalars in a full 2T theory, such as a Higgs
boson doublet $H\left(  X\right)  $ in the Standard Model. But as required by
2T field theory, in particular there is also a flavor-color singlet dilaton
$\Phi\left(  X\right)  $ that gets reduced to $\phi\left(  x\right)  $ and
which must couple as in Eqs.(\ref{redAction1}-\ref{redAction3}). This coupling
of the dilaton disappears in $d=4,$ but there can be additional couplings
among the scalars, such as Higgs and dilaton, which can play a crucial role in
driving the electroweak phase transition that generates masses, by linking it
to other dilaton driven phase transitions, as explained in \cite{2T SM}.}
discussed in \cite{paper1} produced exactly the same overall normalization $Z$
and gave the action for the conformal scalar $\phi\left(  x\right)  $ in the
same background metric $g_{\mu\nu}\left(  x\right)  ,$. The conformal scalar
action is given in Eq.(\ref{confScalar}).

The resulting reduced action $S\left(  A,\Phi\right)  _{red} $ is the action
for a spin-1 gauge field in a variety of shadow curved spacetimes, all with
conformally flat metrics of Eq.(\ref{metric1}). Note that these shadows of the
same $\left(  d+2\right)  $ theory change as the functions $\sigma\left(
x\right)  ,q^{m}\left(  x\right)  $ are arbitrarily chosen. Hence these 1T
field theories must be dual to each other and they must describe the same
gauge invariants from the point of view of $d+2$ dimensions. The duality
transformations among such 1T field theories will be discussed in the next section.

\subsection{Spin-$\frac{1}{2}$ field}

The 2T spin-$\frac{1}{2}$ action, including the gauge field $S\left(
\Psi,A\right)  $ is obtained from (\ref{spinor action}) as usual by replacing
$\partial_{M}$ by the covariant derivative $D_{M}=\partial_{M}-igA_{M}.$
However, in the axial gauge (\ref{axialGauge}) the gauge field drops out in
the kinematic equation (\ref{kinematics}) since $X\cdot D=X\cdot\partial.$
Hence, just like the cases of the scalar and Yang-Mills fields, the spinor
kinematic equation is free from interactions, and simplifies greatly in the
parametrization of Eq.(\ref{parametrization}). So, it takes the form
\begin{equation}
\left(  X\cdot\partial+\frac{d}{2}\right)  \Psi\left(  X\right)  =\left(
\kappa\frac{\partial}{\partial\kappa}+\frac{d}{2}\right)  \Psi\left(
\kappa,\lambda,x^{\mu}\right)  =0,
\end{equation}
which is solved generally by a homogeneous $\Psi$ of degree $-\frac{d}{2}$%
\begin{equation}
\Psi\left(  \kappa,\lambda,x^{\mu}\right)  =\kappa^{-\frac{d}{2}}\hat{\Psi
}\left(  \lambda,x^{\mu}\right)  .
\end{equation}
Expanding $\Psi$ in the form $\Psi=\Psi_{0}+X^{2}\tilde{\Psi}$, we can write
$\hat{\Psi}\left(  \lambda,x^{\mu}\right)  =\Psi_{0}\left(  x^{\mu}\right)
+\left(  \lambda-\frac{q^{2}}{2}\right)  \tilde{\Psi}\left(  \lambda,x^{\mu
}\right)  .$ Using the $X^{2}\zeta_{1}$ part of the 2T gauge symmetry
(\ref{spin half gauge sym}), we can choose the gauge that eliminates the
remainder $\tilde{\Psi}\left(  \lambda,x^{\mu}\right)  =0$, leading to a
$\lambda$-independent $\hat{\Psi}\left(  \lambda,x^{\mu}\right)  =\Psi
_{0}\left(  x^{\mu}\right)  $. Therefore, $\Psi\left(  X\right)  $ takes the
gauge fixed form $\Psi\left(  \kappa,\lambda,x^{\mu}\right)  =\kappa
^{-\frac{d}{2}}\Psi_{0}\left(  x^{\mu}\right)  .$ Using now the (kappa type)
$\not X  \zeta_{2}$ part of the gauge symmetry (\ref{spin half gauge sym}), we
can remove half of the remaining degrees of freedom of $\Psi_{0}$. \ In
particular, one can make the gauge choice
\begin{equation}
\Gamma^{+^{\prime}}\Psi=0. \label{gaugedPs}%
\end{equation}
\ With a choice of basis for the flat space SO$\left(  d,2\right)  $ gamma
matrices $\Gamma^{M}$ (see Appendix of \cite{super2t}), the gauge condition
$\Gamma^{+^{\prime}}\Psi=0$ forces the lower components of $\Psi$ (equivalent
to the first two components of $\bar{\Psi}$) to vanish. Therefore, the gauge
fixed form of $\Psi\left(  X\right)  $ is%
\begin{equation}
\Psi\left(  \kappa,\lambda,x^{\mu}\right)  =\kappa^{-\frac{d}{2}}%
e^{-\frac{\sigma\left(  x\right)  }{2}}\left(
\begin{array}
[c]{c}%
\psi\left(  x^{\mu}\right) \\
0
\end{array}
\right)  ,\;\bar{\Psi}\left(  \kappa,\lambda,x^{\mu}\right)  =\kappa
^{-\frac{d}{2}}e^{-\frac{\sigma\left(  x\right)  }{2}}\left(
\begin{array}
[c]{cc}%
0 & \bar{\psi}\left(  x^{\mu}\right)
\end{array}
\right)  \label{gaugedPsi}%
\end{equation}
where $\psi\left(  x^{\mu}\right)  $ is an SO$\left(  d-1,1\right)  $ spinor
and $\bar{\psi}\left(  x^{\mu}\right)  $ is its anti-spinor. We have inserted
the extra factor $e^{-\frac{\sigma}{2}}$ for later convenience in the
interpretation of $\psi.$

We now focus on the term $\left(  \bar{\Psi}\not X  \not \bar {D}\Psi\right)
$ in the action, where $\not \bar {D}=\bar{\Gamma}^{M}D_{M}=\bar{\Gamma
}^{+^{\prime}}D_{+^{\prime}}+\bar{\Gamma}^{-^{\prime}}D_{-^{\prime}}%
+\bar{\Gamma}^{m}D_{m}$ includes the gauge field. \ With our gauge choices in
Eqs.(\ref{gaugedA},\ref{gaugedPs},\ref{gaugedPsi}) we can drop the terms,
$\bar{\Gamma}^{+^{\prime}}D_{+^{\prime}}\Psi=0$ and $\bar{\Gamma}^{-^{\prime}%
}D_{-^{\prime}}\Psi=\bar{\Gamma}^{-^{\prime}}\frac{1}{\kappa}e^{-\sigma}%
\frac{\partial}{\partial\lambda}\Psi=0$, and using explicitly our SO$\left(
d,2\right)  $ gamma matrices we get%
\begin{equation}
\not \bar {D}\Psi=\bar{\Gamma}^{m}D_{m}\Psi=\left(
\begin{array}
[c]{cc}%
\gamma^{m}D_{m} & 0\\
0 & -\bar{\gamma}^{m}D_{m}%
\end{array}
\right)  \left(
\begin{array}
[c]{c}%
\kappa^{-\frac{d}{2}}e^{-\frac{\sigma}{2}}\psi\\
0
\end{array}
\right)  =\left(
\begin{array}
[c]{c}%
\gamma^{m}D_{m}\left(  \kappa^{-\frac{d}{2}}e^{-\frac{\sigma}{2}}\psi\right)
\\
0
\end{array}
\right)
\end{equation}
where $\gamma^{m}$ are now the SO$\left(  d-1,1\right)  $ gamma matrices in
\textit{flat tangent space }labeled by\textit{ }$m$. Next,$\ $we apply
$\not X  =\Gamma^{M}X_{M}=-\Gamma^{+^{\prime}}X^{-^{\prime}}-\Gamma
^{-^{\prime}}X^{+^{\prime}}+\Gamma^{m}X_{m}$ on $\not \bar {D}\Psi.$ \ The
first term $\Gamma^{+^{\prime}}X^{-^{\prime}}$ gives zero when acting on
$\not \bar {D}\Psi$. \ The other two terms give%
\begin{align}
\not X  \not \bar {D}\Psi &  =\kappa e^{\sigma}\left(
\begin{array}
[c]{cc}%
\bar{\gamma}^{m}q_{m} & 0\\
1 & -\gamma^{m}q_{m}%
\end{array}
\right)  \left(
\begin{array}
[c]{c}%
\gamma^{k}D_{m}\left(  \kappa^{-\frac{d}{2}}e^{-\frac{\sigma}{2}}\psi\right)
\\
0
\end{array}
\right) \nonumber\\
&  =\kappa e^{\sigma}\left(
\begin{array}
[c]{c}%
\bar{\gamma}^{m}q_{m}\gamma^{k}D_{k}\left(  \kappa^{-\frac{d}{2}}%
e^{-\frac{\sigma}{2}}\psi\right) \\
\gamma^{k}D_{k}\left(  \kappa^{-\frac{d}{2}}e^{-\frac{\sigma}{2}}\psi\right)
\end{array}
\right)  .
\end{align}
Therefore%
\begin{align}
\bar{\Psi}\not X  \not \bar {D}\Psi &  =\kappa e^{\sigma}e^{-\frac{\sigma}{2}%
}\kappa^{-\frac{d}{2}}\left(
\begin{array}
[c]{cc}%
0 & \bar{\psi}%
\end{array}
\right)  \left(
\begin{array}
[c]{c}%
\bar{\gamma}^{m}q_{m}\gamma^{k}D_{k}\left(  \kappa^{-\frac{d}{2}}%
e^{-\frac{\sigma}{2}}\psi\right) \\
\gamma^{k}D_{k}\left(  \kappa^{-\frac{d}{2}}e^{-\frac{\sigma}{2}}\psi\right)
\end{array}
\right) \nonumber\\
&  =\kappa^{-d}\bar{\psi}\gamma^{k}e_{k}^{\mu}\left(  D_{\mu}+\frac{d-1}%
{2}\partial_{\mu}\sigma\right)  \psi. \label{result}%
\end{align}

As an additional step, we note the identity
\begin{equation}
\frac{d-1}{2}\left(  \gamma^{k}e_{k}^{\mu}\right)  \partial_{\mu}\sigma
=\frac{1}{4}\left(  \gamma^{k}\gamma_{ij}\right)  e_{k}^{\mu}\omega_{\mu}^{ij}%
\end{equation}
that expresses the term that contains $\partial_{\mu}\sigma$ as due to the
following special spin connection $\omega_{\mu}^{ij}\left(  x\right)  $ for
the SO$\left(  d-1,1\right)  $ in tangent space
\begin{equation}
\omega_{\mu}^{ij}\left(  x\right)  =\left(  e_{\mu}^{i}e^{j\nu}-e_{\mu}%
^{j}e^{i\nu}\right)  \partial_{\nu}\sigma\left(  x\right)  . \label{spinConn1}%
\end{equation}
But this $\omega_{\mu}^{ij}\left(  x\right)  $ is precisely the spin
connection that is constructed from the vielbein in curved space as usual%
\begin{equation}
\omega_{\mu}^{ij}\left(  x\right)  =e^{i\lambda}e^{j\sigma}\left(
c_{\mu\lambda\sigma}-c_{\lambda\sigma\mu}-c_{\sigma\mu\lambda}\right)  ,\text{
~with~}c_{\mu\lambda\sigma}\equiv-\frac{1}{2}e_{\mu}^{k}\left(  \partial
_{\lambda}e_{\sigma k}-\partial_{\sigma}e_{\lambda k}\right)  .
\label{spinConn2}%
\end{equation}
If we insert our $e_{\mu}^{i}=e^{\sigma}\partial_{\mu}q^{i}\left(  x\right)  $
in this general expression, we recover precisely the special spin connection
above. Therefore, the result in Eq.(\ref{result}) can now be written as%
\begin{equation}
(\bar{\Psi}\not X  \not \bar {D}\Psi)_{red}=\kappa^{-d}\bar{\psi}\gamma
^{k}e_{k}^{\mu}\hat{D}_{\mu}\psi
\end{equation}
where the covariant derivative $\hat{D}_{\mu}$ includes both the Yang-Mills
and the spin connection in $\left(  d-1\right)  +1$ dimensions.%
\begin{equation}
\hat{D}_{\mu}=\partial_{\mu}-igA_{\mu}+\frac{1}{4}\omega_{\mu}^{ij}\gamma
_{ij}.
\end{equation}

The reduced action in which the kinematic constraints are solved, now takes
the form%
\begin{align}
S\left(  \Psi,A\right)  _{red}  &  =\frac{i}{2}Z\int\left(  d^{d+2}X\right)
\delta\left(  X^{2}\right)  \left(  \bar{\Psi}\not X  \not \bar {D}%
\Psi+h.c.\right)  _{red}\nonumber\\
&  =\frac{i}{2}Z\int d\kappa d\lambda d^{d}x\frac{1}{2}\kappa^{d-1}\det\left(
e_{\mu}^{m}\right)  ~\delta\left(  \lambda-\frac{q^{2}\left(  x\right)  }%
{2}\right)  \left(  \kappa^{-d}\bar{\psi}\gamma^{k}e_{k}^{\mu}\hat{D}_{\mu
}\psi+h.c.\right) \nonumber\\
&  =\frac{1}{2}\int d^{d}x\sqrt{-g}\bar{\psi}i\gamma^{k}\hat{D}_{k}\psi+h.c
\label{reducedSpinor}%
\end{align}
where we have used again the volume element in (\ref{volume}) as well as the
previous universal normalization $Z\int\frac{d\kappa}{2\kappa}=1$.

As in the cases of the scalar and vector fields, the resulting spinor action
in Eq.(\ref{reducedSpinor}) is the standard 1T field theory action in a
$(d-1)+1$ curved spacetime. The conformally flat metric $g^{\mu\nu}\left(
x\right)  $ is again the same one that describes the shadow spacetime for the
other fields. As before, the shadows are different as we change the functions
$\sigma\left(  x^{\mu}\right)  $ and $q^{m}\left(  x^{\mu}\right)  .$ So the
field theories with the different conformally flat backgrounds must be dual to
each other since each shadow must describe the same gauge invariant content of
the original 2T field theory in $d+2$ dimensions.

\section{Dualities}

We have shown above that the 2T field theory in $d+2$ dimensions leads to a
family of 1T field theories corresponding to all possible conformally flat
backgrounds in $\left(  d-1\right)  +1$ dimensions. \ We now show the
relations that transform one shadow with a fixed spacetime metric $g_{\mu\nu
}\left(  x\right)  $ (example, flat Minkowski spacetime) into another shadow
with a different spacetime metric $\tilde{g}_{\mu\nu}$ (example,
Robertson-Walker expanding universe). From the point of view of 1T physics,
this is a transformation between two different theories with no a priori
relation to each other. But from the point of view of 2T physics, from the
derivation above, it is evident that such transformations among 1T field
theories should be an actual symmetry among the shadows that does not change
the physical content, and hence we call it a duality transformation in 1T-physics.\

The duality transformations that we will discuss here take the following form%
\begin{equation}
S_{\sigma,q_{m}}\left(  \phi,A_{\mu},\psi\right)  =S_{\tilde{\sigma},\tilde
{q}_{m}}\left(  \tilde{\phi},\tilde{A}_{\mu},\tilde{\psi}\right)
\label{duality}%
\end{equation}
On the left side $S_{\sigma,q_{m}}\left(  \phi,A_{\mu},\psi\right)  $
represents the 1T field theory with scalars,
vectors and spinors in a background geometry generated by the
functions $\sigma\left(  x\right) ,q_{m}\left(
x\right)  .$ On the right side the background geometry has been
changed to a new one $\tilde{\sigma}\left(
x\right)  ,\tilde{q}_{m}\left( x\right)  ,$ and when the dynamical
fields are transformed into new ones by a
duality transformation $\left(  \phi,A_{\mu},\psi\right)  \rightarrow\left(
\tilde{\phi},\tilde{A}_{\mu},\tilde{\psi}\right)  $, the actions can
be shown to be equal. Hence, such a duality
transformation is a symmetry of the system. Of course, this symmetry
among the shadows is a simple consequence of
the fact that either expression is merely a parametrization of the
solutions of the kinematic constraints
(\ref{kinematics}) of the same 2T action
\begin{equation}
S_{\sigma,q_{m}}\left(  \phi,A_{\mu},\psi\right)  =S\left(  \Phi,A_{M}%
,\Psi_{\alpha}\right)  _{red}=S_{\tilde{\sigma},\tilde{q}_{m}}\left(
\tilde{\phi},\tilde{A}_{\mu},\tilde{\psi}\right)  .
\end{equation}
In 1T physics we now verify directly that the sample cases given in Table-1,
indeed form a set of dual field theories.

We consider the following two types of \textit{local} transformations of the
background functions that relate a subset of the shadow spacetimes to one another.

\begin{itemize}
\item First consider replacing the functions $q^{n}\left(  x\right)  $ by new
ones $\tilde{q}^{m}\left(  x\right)  .$ This can be implemented by general
coordinate transformation in $q$-space $q^{m}\rightarrow\tilde{q}^{m}\left(
q\right)  ,$ which yields $\tilde{q}^{m}\left(  x\right)  =\tilde{q}%
^{m}\left(  q\left(  x\right)  \right)  .$ Since general coordinate
transformations in $x$-space $x^{\mu}\rightarrow y^{\mu}\left(  x\right)  $
have the same amount of freedom as $q$-space reparametrizations, the resulting
function $\tilde{q}^{m}\left(  x\right)  $ can also be built through general
$x$-reparametrizations. Thus we can write $\tilde{q}^{m}\left(  x\right)  $ in
two ways $\tilde{q}^{m}\left(  q^{n}\left(  x\right)  \right)  =\tilde{q}%
^{m}\left(  x\right)  =q^{m}\left(  y\left(  x\right)  \right)  $. To prove
the duality in Eq. (\ref{duality}) we will treat $q$-reparametrization as
general coordinate transformations in $x$-space. In that case the background
functions $\sigma\left(  x\right)  $ and $q^{m}\left(  x\right)  $ are
transformed like general coordinate scalars%
\begin{equation}
\tilde{\sigma}\left(  x\right)  =\sigma\left(  y\left(  x\right)  \right)
,\;\;\tilde{q}^{m}\left(  x\right)  =q^{m}\left(  y\left(  x\right)  \right)
. \label{genCoord}%
\end{equation}
These induce general coordinate transformations on the background geometry
$e_{\mu}^{m}\left(  x\right)  \rightarrow\tilde{e}_{\mu}^{m}\left(  x\right)
=e^{\tilde{\sigma}\left(  x\right)  }\partial_{\mu}\tilde{q}^{m}\left(
x\right)  $ which takes the form
\begin{equation}
\tilde{e}_{\mu}^{m}\left(  x\right)  =\partial_{\mu}y^{\lambda}\left(
x\right)  e_{\lambda}^{m}\left(  y\left(  x\right)  \right)  ,\;\;\tilde
{g}_{\mu\nu}\left(  x\right)  =\partial_{\mu}y^{\lambda}\left(  x\right)
\partial_{\nu}y^{\sigma}\left(  x\right)  g_{\lambda\sigma}\left(  y\left(
x\right)  \right)  . \label{genCoord3}%
\end{equation}

\item Now consider changing $\sigma\left(  x\right)  $ to a new one, leaving
$q^{m}\left(  x\right)  $ alone. This can be implemented as follows%
\begin{equation}
\tilde{\sigma}\left(  x\right)  =\sigma\left(  x\right)  +\lambda\left(
x\right)  ,\;\;\tilde{q}^{m}\left(  x\right)  =q^{m}\left(  x\right)
\label{weyl}%
\end{equation}
This induces a scale transformation on both the vielbein and metric%
\begin{equation}
\tilde{e}_{\mu}^{m}\left(  x\right)  =e^{\lambda\left(  x\right)  }e_{\mu}%
^{n}\left(  x\right)  ,\;\;\tilde{g}_{\mu\nu}\left(  x\right)  =e^{2\lambda
\left(  x\right)  }g_{\mu\nu}\left(  x\right)  . \label{Weyl transf metric}%
\end{equation}
Hence the change $\sigma\left(  x\right)  \rightarrow\tilde{\sigma}\left(
x\right)  $ amounts to a Weyl transformation.
\end{itemize}

Since the reduced 1T action is formally invariant under general coordinate
transformations, we can claim that the action with background $\left(
\sigma,q_{m}\right)  \left(  x\right)  $ will be equal to the action with
background $\left(  \tilde{\sigma},\tilde{q}_{m}\right)  \left(  y\left(
x\right)  \right)  $ as in Eq.(\ref{duality}) provided the fields $\left(
\phi,A_{\mu},\psi\right)  $ are also transformed by the general coordinate
transformations%
\begin{equation}
\tilde{\phi}\left(  x\right)  =\phi\left(  y\left(  x\right)  \right)
,\;\tilde{A}_{\mu}\left(  x\right)  =\partial_{\mu}y^{\lambda}\left(
x\right)  A_{\lambda}\left(  y\left(  x\right)  \right)  ,\;\tilde{\psi
}\left(  x\right)  =\psi\left(  y\left(  x\right)  \right)  \label{genCoord2}%
\end{equation}
This is then the duality transformation that relates actions with the two
different backgrounds in Eq.(\ref{genCoord}).

A less obvious duality symmetry is Weyl transformations given by the
transformation of the background geometry in Eqs.(\ref{weyl}%
,\ref{Weyl transf metric}) and the following transformations of the dynamical
fields
\begin{equation}
\tilde{\phi}\left(  x\right)  =e^{-\frac{d-2}{2}\lambda\left(  x\right)  }%
\phi\left(  x\right)  ,\;\;\tilde{A}_{\mu}\left(  x\right)  =A_{\mu}\left(
x\right)  ,\;\;\tilde{\psi}\left(  x\right)  =e^{-\frac{d-1}{2}\lambda\left(
x\right)  }\psi\left(  x\right)  . \label{weyl2}%
\end{equation}
We will now prove that this is a duality symmetry as in Eq.(\ref{duality}).

For the spin-1 action in Eq.(\ref{redAction3}), note that%
\begin{equation}
\tilde{F}_{\mu\nu}\tilde{F}^{\mu\nu}=F_{\mu\nu}\tilde{g}^{\mu\lambda}\tilde
{g}^{\nu\sigma}F_{\lambda\sigma}=e^{-4\lambda}F_{\mu\nu}g^{\mu\lambda}%
g^{\nu\sigma}F_{\lambda\sigma}=e^{-4\lambda}F_{\mu\nu}F^{\mu\nu}%
\end{equation}
Then, the transformed action is seen to be invariant%
\begin{align}
\tilde{S}\left(  \tilde{A},\tilde{\phi}\right)   &  =\int d^{d}x\left(
e^{\lambda d}\sqrt{-g}\right)  \left(  -\frac{1}{4}\left(  e^{-\frac{d-2}%
{2}\lambda}\phi\right)  ^{\frac{2\left(  d-4\right)  }{d-2}}Tr\left(
F_{\mu\nu}e^{-4\lambda}F^{\mu\nu}\right)  \right) \\
&  =\int d^{d}x~e^{\left(  d-\left(  d-4\right)  -4\right)  \lambda}\sqrt
{-g}\left(  -\frac{1}{4}\phi^{\frac{2\left(  d-4\right)  }{d-2}}Tr\left(
F_{\mu\nu}F^{\mu\nu}\right)  \right) \\
&  =S\left(  A,\phi\right)
\end{align}
which proves the duality symmetry when the background and dynamical fields are
transformed according to (\ref{genCoord},\ref{genCoord2}).

For the spin-$\frac{1}{2}$ case in Eq.(\ref{reducedSpinor}), it is faster to
prove the duality if we use the version of the covariant derivative in
Eq.(\ref{result}). Then we have%
\begin{align}
\tilde{S}\left(  \tilde{\psi},\tilde{A}\right)   &  =\frac{1}{2}\int
d^{d}x\left\{
\begin{array}
[c]{c}%
\left(  e^{\lambda d}\sqrt{-g}\right)  \left(  e^{-\frac{d-1}{2}\lambda}%
\bar{\psi}\right)  i\gamma^{k}\left(  e^{-\lambda}e_{k}^{\mu}\right) \\
\times\left(  D_{\mu}+\frac{d-1}{2}\partial_{\mu}\left(  \sigma+\lambda
\right)  \right)  \left(  e^{-\frac{d-1}{2}\lambda}\psi\right)
\end{array}
\right\}  +h.c.\\
&  =\frac{1}{2}\int d^{d}x\left\{
\begin{array}
[c]{c}%
e^{\lambda d}e^{-\frac{d-1}{2}\lambda}e^{-\lambda}e^{-\frac{d-1}{2}\lambda
}\sqrt{-g}\bar{\psi}i\gamma^{k}e_{k}^{\mu}\\
\times\left(  D_{\mu}-\frac{d-1}{2}\partial_{\mu}\lambda+\frac{d-1}{2}%
\partial_{\mu}\left(  \sigma+\lambda\right)  \right)  \psi
\end{array}
\right\}  +h.c.\\
&  =S\left(  \psi,A\right)
\end{align}
So it is invariant under Weyl transformations, which proves the duality
symmetry when the background and dynamical fields are transformed according to
(\ref{weyl},\ref{weyl2}).

Dualities under general coordinate transformations and Weyl transformations of
the type above hold for all background metrics $g_{\mu\nu}$, not only for the
conformally flat metrics, so is there something more special in the present
case, and how would the general case be recovered in 2T-physics? The answer is
found by recalling that we have investigated duality properties of the shadows
of a specific 2T-theory.

First, we must emphasize that there are more shadows of the same theory that
are not conformally flat field theories, but also participate in similar
duality transformations. Those have not been discussed in our preliminary work
in this paper as explained in footnote (\ref{others}), as our main motivation
here was to provide some simple examples of the dualities generated by
2T-field theory.

Second, the starting point can be various 2T-field theories, including curved
backgrounds in $d+2$ dimensions rather than the flat background in
(\ref{flat}) used in our present case. Curved backgrounds in $d+2$ dimensions
will lead to shadows in more general backgrounds $g_{\mu\nu}$ that would not
be necessarily conformally flat, but will satisfy the dualities generated by
Weyl and general coordinate transformations as in the more general case.

Third, by starting from a specific 2T-theory we can generate only those
shadows that capture the underlying properties of that theory. So, the
conformally flat spacetimes represented by a subset of shadows in Table-1,
must have additional properties that reflect the properties of flat spacetime
in $d+2$ dimensions. Specifically, these shadows must have a hidden SO$\left(
d,2\right)  $ \textit{global} symmetry. This additional property of each
shadow is discussed in the next section.

\section{SO$\left(  d,2\right)  $ global symmetry and its generators
\label{globalsymm}}

Similarly to the spin-0 case, the SO$\left(  d,2\right)  $ global symmetry of
the original 2T theory must still be present after imposing the SO$\left(
d,2\right)  $ invariant kinematic constraints (\ref{kinematics}). In this
section, we provide the explicit form of the SO$\left(  d,2\right)  $
generators $J_{MN}$ as applied on each shadow.

To do so, we will use the same trick as in \cite{paper1}. \ The generic field
$\chi_{\mu_{1}\mu_{2}\cdots}$ is a shadow in curved space, with metric
\begin{equation}
g_{\mu\nu}\left(  x\right)  =e^{2\sigma\left(  x\right)  }\eta_{mn}%
\partial_{\mu}q^{m}\left(  x\right)  \partial_{\nu}q^{n}\left(  x\right)  .
\label{genShadow}%
\end{equation}
This can be related by dualities to the shadow field $\chi_{\mu_{1}\mu
_{2}\cdots}^{0}$, in flat space with metric $\eta^{\mu\nu}.$ The duality
relation is a combination of Weyl and general coordinate transformations
parametrized by $\lambda\left(  x\right)  $ and $y^{\mu}\left(  x\right)  $ as
shown in Eqs.(\ref{genCoord},\ref{genCoord2},\ref{weyl},\ref{weyl2}). The
starting point we want to use for the duality transformation is the shadow
with flat space Minkowski metric $\eta_{\mu\nu}$, hence we want to apply the
combined transformations in Eqs.(\ref{genCoord},\ref{weyl}) as follows
\begin{equation}
\eta_{\mu\nu}\rightarrow g_{\mu\nu}\left(  x\right)  =e^{2\lambda\left(
x\right)  }\partial_{\mu}y^{\rho}\left(  x\right)  \partial_{\nu}y^{\sigma
}\left(  x\right)  \eta_{\rho\sigma}.
\end{equation}
Then, the parameters $\lambda\left(  x\right)  ,y^{\mu}\left(  x\right)  $
that produce the general shadow spacetime of Eq.(\ref{genShadow}) are
precisely $\lambda\left(  x\right)  =\sigma\left(  x\right)  $ and $y^{\mu
}\left(  x\right)  =\delta_{m}^{\mu}q^{m}\left(  x\right)  .$ Hence by these
duality transformations, the generic field $\chi$ can be written in terms of
the flat-space field $\chi^{0}$ as in Eqs.(\ref{genCoord2},\ref{weyl2})
\begin{equation}
\chi_{\mu_{1}\mu_{2}\cdots}\left(  x\right)  =e^{-w\sigma\left(  x\right)
}\left(  \frac{\partial y^{\nu_{1}}}{\partial x^{\mu_{1}}}\frac{\partial
y^{\nu_{2}}}{\partial x^{\mu_{2}}}\cdots\right)  \chi_{\nu_{1}\nu_{2}\cdots
}^{0}\left(  y\left(  x\right)  \right)  , \label{duality2}%
\end{equation}
where $w$ is the Weyl weight of the field $\chi$ as it appears in
Eq.(\ref{weyl2}) for the relevant fields in this paper.

Now we begin to investigate the SO$\left(  d,2\right)  $ transformations. The
starting point is the form of the SO$\left(  d,2\right)  $ generators in the
$d+2$ theory, which is
\begin{equation}
J^{MN}=\left(  X^{M}P^{N}-X^{N}P^{M}\right)  +S^{MN},
\end{equation}
where $P_{M}=$ $-i\partial/\partial X^{M}$ is a differential operator as
applied on any 2T field, and $S^{MN}$ is the the representation of SO$\left(
d,2\right)  $ as applied on the spin indices of the fields%
\begin{equation}
S^{MN}\Phi=0,\;S^{MN}A_{K}=\left(  \delta_{K}^{M}\eta^{NL}-\delta_{K}^{N}%
\eta^{ML}\right)  A_{L},\;S^{MN}\Psi_{\alpha}=\frac{1}{2i}\left(  \Gamma
^{MN}\right)  _{\alpha}^{~\beta}\Psi_{\beta}.
\end{equation}

Note that the kinematic conditions (\ref{kinematics}) are invariant under
these transformations. Therefore, this form of $J^{MN}$ implies corresponding
transformations for the shadow fields in the lower dimension. For a particular
parametrization, such as Eq.(\ref{parametrization}) for $X^{M}$ and
Eqs.(\ref{chain1}-\ref{chain3}) for $P_{M}=$ $-i\partial/\partial X^{M}$ , the
generator $J^{MN}$ implements the infinitesimal SO$\left(  d,2\right)  $
transformation on the original fields $\Phi,A_{M},\Psi_{\alpha}$ but now as
functions of $\kappa,\lambda,x^{\mu}.$ When we insert the solutions of the
kinematic equations, all $\partial/\partial\lambda$ derivatives vanish on the
solutions, any remaining explicit $\lambda$ is replaced by $\lambda
=q^{2}\left(  x\right)  /2$, and the $\kappa$ dependence becomes trivial since
it appears as overall factors$.$ In particular, if we pick the parametrization
that corresponds to shadows in Minkowski space (i.e. $\sigma=0$ and
$q^{m}\left(  x\right)  =\delta_{\mu}^{m}x^{\mu}$ in Eq.(\ref{parametrization}%
)), then the shadow fields $\chi^{0}\left(  q\right)  $ transform under
SO$\left(  d,2\right)  $ as
\begin{equation}
\text{flat:\ }\delta_{\omega}\chi^{0}\left(  q\right)  =\frac{i}{2}\omega
_{MN}J_{0}^{MN}\chi^{0}.
\end{equation}
where $J_{0}^{MN}$ takes the form of the familiar conformal transformations
\cite{2T basics} (contrast to the classical version in Eqs.(\ref{jmn1}%
-\ref{jmn3}) when specialized to $\sigma=0$ and $q^{m}\left(  x\right)
=\delta_{\mu}^{m}x^{\mu}$)
\begin{equation}
J_{0}^{mn}=\left(  q^{m}p^{n}-q^{n}p^{m}\right)  +S^{mn}~\text{(Lorentz
transf.)}%
\end{equation}%
\begin{equation}
J_{0}^{+^{\prime}m}=p^{m}~\text{(translations);\ \ }J_{0}^{+^{\prime}%
-^{\prime}}=q^{m}p_{m}-ik~\text{(dilatations)}%
\end{equation}%
\begin{equation}
J_{0}^{-^{\prime}m}=\frac{1}{2}q_{l}q^{l}p^{m}-q^{m}q^{l}p_{l}-q_{l}%
S^{ml}+ikq^{m}~\text{(conformal transf.)}%
\end{equation}
where $p_{m}$\ is understood as a differential operator
\begin{equation}
p_{m}\equiv-i\frac{\partial}{\partial q^{m}},
\end{equation}
$S^{mn}$ is the appropriate spinor representation of SO$\left(  d-1,1\right)
$, and $k$ is the scaling dimension of the corresponding field%
\begin{equation}
k_{\phi}=\frac{d-2}{2},\;\;k_{\psi}=\frac{d-1}{2},\;\;k_{A}=1.
\end{equation}

Now, by using the duality transformation (\ref{duality2}), we derive the
SO$\left(  d,2\right)  $ transformation for the fields in the curved
background as $\delta_{\omega}\chi_{i}\left(  x\right)  =e^{-w\sigma}%
\Lambda_{i}^{\text{ \ }j}\left(  \delta_{\omega}\chi_{j}^{0}\left(  q\left(
x\right)  \right)  \right)  $. We obtain (here the indices $i$ and the symbol
$\Lambda_{i}^{\text{ \ }j}$ is short hand notation for those that appear in
Eq.(\ref{duality2}))%
\begin{align}
\delta_{\omega}\chi_{i}\left(  x\right)   &  =e^{-w\sigma}\Lambda_{i}^{\text{
\ }j}\left(  \delta_{\omega}\chi_{j}^{0}\left(  q\left(  x\right)  \right)
\right)  =\frac{i}{2}\omega_{MN}~e^{-w\sigma}\Lambda_{i}^{\text{ \ }j}%
J_{0}^{MN}\chi_{j}^{0}\left(  q\left(  x\right)  \right) \\
&  =\frac{i}{2}\omega_{MN}~e^{-w\sigma}\Lambda_{i}^{\text{ \ }j}J_{0}%
^{MN}\left(  e^{w\sigma}\left(  \Lambda^{-1}\right)  _{j}^{\text{ \ }k}%
\chi_{k}\left(  x\right)  \right)  \equiv\frac{i}{2}\omega_{MN}~J^{MN}%
~\chi_{i}\left(  x\right)
\end{align}
Hence the action of the $J^{MN}$ defined by the last expression is given by
the differential operators%
\begin{equation}
J^{MN}~\chi_{i}\left(  x\right)  =e^{-w\sigma}\Lambda_{i}^{\text{ \ }j}%
J_{0}^{MN}\left[  e^{w\sigma}\left(  \Lambda^{-1}\right)  _{j}^{\text{ \ }%
k}\chi_{k}\left(  x\right)  \right]  \label{LMN}%
\end{equation}

Let us now specialize to the cases of spin-$\frac{1}{2}$ and spin-1 fields
(spin-0 is given in \cite{paper1}). Since $\psi\left(  x\right)
=e^{-\frac{d-1}{2}\sigma}\psi_{\beta}^{\left(  0\right)  }\left(  q\left(
x\right)  \right)  ,$ the SO$\left(  d,2\right)  $ generators for spin $1/2$
fields in conformally flat curved space are given by%
\begin{equation}
J^{MN}\psi\left(  x\right)  =\left[  e^{-\frac{d-1}{2}\sigma\left(  x\right)
}J_{0}^{MN}e^{\frac{d-1}{2}\sigma\left(  x\right)  }\right]  \psi\left(
x\right)
\end{equation}
Similarly, since $A_{\mu}\left(  x\right)  =e_{\mu}^{m}\left(  x\right)
A_{m}^{\left(  0\right)  }\left(  q\left(  x\right)  \right)  ,$ the
SO$\left(  d,2\right)  $ generators for spin $1$ fields in conformally flat
curved space are given by%
\begin{equation}
J^{MN}A_{\mu}\left(  x\right)  =\left[  e_{\mu}^{i}\left(  x\right)
J_{0}^{MN}e_{i}^{\nu}\left(  x\right)  \right]  A_{\nu}\left(  x\right)  .
\end{equation}
In these expressions to compute the action of $p_{m}$ that appears in
$J_{0}^{MN}$ we just use the chain rule to apply $p_{m}$ on any function of
$x$ as follows
\begin{equation}
p_{m}f\left(  x\right)  =-i\frac{\partial}{\partial q^{m}}f\left(  x\right)
=-i\frac{\partial x^{\mu}}{\partial q^{m}}\frac{\partial}{\partial x^{\mu}%
}f\left(  x\right)  =-ie^{\sigma}e_{m}^{\mu}\frac{\partial}{\partial x^{\mu}%
}f\left(  x\right)  ,
\end{equation}
where we inserted $\frac{\partial x^{\mu}}{\partial q^{m}}=e^{\sigma}%
e_{m}^{\mu}$ as discussed before. Then the resulting expressions are the
quantum ordered versions of the classical generators given in Eqs.(\ref{jmn1}%
-\ref{jmn3})

We emphasize that the fixed background metric $g_{\mu\nu}\left(  x\right)  $
is unchanged by the global SO$\left(  d,2\right)  $ transformation (this is
seen easily from the construction of $g_{\mu\nu}$ in Eq.(\ref{flat2})).
Therefore, without reference to the flat theory, but only using the generator
$J^{MN}$ above, we see that this is a true invariance of the action with the
fixed background $\left(  \sigma,q^{i}\right)  $%
\begin{equation}
\delta_{\omega}S_{\sigma,\chi}\left(  \phi,A_{\mu},\psi_{\alpha}\right)
=\frac{\partial S_{\sigma,q}}{\partial\phi}\delta_{\omega}\phi+\frac{\partial
S_{\sigma,q}}{\partial A_{\mu}}\delta_{\omega}A_{\mu}+\frac{\partial
S_{\sigma,q}}{\partial\psi_{\alpha}}\delta_{\omega}\psi_{\alpha}=0.
\label{delSfixed}%
\end{equation}
This hidden global symmetry is nothing but the original global SO$\left(
d,2\right)  $ symmetry of the action $S\left(  \Phi,A,\Psi\right)  $ in $d+2$
dimensions, and hence each shadow for any $\left(  \sigma\left(  x\right)
,q^{i}\left(  x\right)  \right)  $ must also be invariant.

It is straightforward to see that there is a symmetry as in
Eq.(\ref{delSfixed}) for each shadow, when presented as an outcome of the
higher dimensional formulation, but this symmetry is not so easy to spot for
specific backgrounds in 1T-physics field theory. For example, we claim that
the emergent field theory in the Robertson-Walker expanding universe has this
hidden SO$\left(  d,2\right)  $ global symmetry, which was not noticed before.
The resulting expressions for the hidden SO$\left(  d,2\right)  $ generators
$J^{MN}$ given above are new.

The Robertson-Walker example, as well as all the others listed in Table-1,
show that 1T-physics is not equipped to predict the hidden symmetries or
dualities. However, within 1T-physics field theory, with hard work and some
guidance, one can find new properties, such as the dualities and hidden
symmetries descibed above. Within 1T-physics these are the clues as well as
the evidence of the higher dimensional nature of the underlying 2T spacetime,
as predicted by 2T-physics.

\section{Conclusion}

In this paper, we have shown that 1T field theories involving Dirac and
Yang-Mills fields propagating in any conformally flat metric in $\left(
d-1\right)  +1$ dimensions can be obtained as the shadows of 2T field theory
in flat $d+2$ dimensions. \ This generalizes a similar result for the
conformal scalar field as reported in \cite{paper1}. Since the shadows belong
to the same parent theory, there has to be hidden relationships among the
emergent 1T field theories. We have displayed some of these hidden
relationships in the form of dualities and also in the form of hidden
symmetries, which were not previously known to exist for many of the specific
examples listed in Table-1.

This, of course can be applied to theories with several interacting fields of
different spins as is the case of the Standard Model. \ Indeed the usual
Standard model in $3+1$ dimensions is already known to be the flat Minkowski
shadow of a corresponding field theory in $4+2$ dimensions \cite{2T SM}, and
therefore our approach in this paper, which extends also to the shadows of the
Standard Model, may find practical applications.

We should emphasize that the particular class of shadow spacetimes that we
have discussed only constitutes a starting point. \ The infinity of possible
gauge choices in the worldline formalism suggests a similar richness in 2T
field theory. \ In particular, we would like to extend the theory to allow
gauge choices equivalent to those in the worldline formalism which involve
mixing of $x$ and $p$ (footnote (\ref{others})). \ This may result in
dualities between local and non-local field theories at least in some
instances. \ It is to be noted that the appearance of mass, coupling and
curvature, as moduli in the worldline formalism was related to such gauge
choices. \ Here we have seen examples of 1T field theory where curvature
emerged as moduli in the reduction from 2T field theory. This suggests the
possibility that mass in field theory might also come as a modulus in the
embedding of 3+1 dimensional phase space into 4+2 dimensional phase space.
This is a topic which is worth pursuing in more detail.

We also believe that the more general dualities provided by 2T-physics could
provide new tools to investigate the properties of the Standard Model,
including QCD. For instance, one could use one form of the 1T-physics action
to learn some non-perturbative information about the other 1T-physics action.
This suggests that we may be able to take advantage of the type of dualities
discussed here, and their extensions (as suggested in footnote (\ref{others}%
)), to develop non-perturbative tools for analyzing the Standard Model itself
as well as its dual versions.

So far, our discussion of field theory was purely classical. \ Another goal of
our program is the quantization of our theory directly\ in the 2T formulation.
This step is obviously necessary in order to fully express the Standard Model
as a 2T theory at a quantum level. This is being pursued in the path integral
formalism, taking into consideration the Faddeev-Popov formalism for gauge
fixing the local symmetries of the 2T-field theory. \

Further research on these topics is warranted and is currently being pursued.

\begin{acknowledgments}
We would like to thank S.-H Chen, Y.-C. Kuo and B. Orcal for helpful discussions.
\end{acknowledgments}

\appendix

\section{Relating worldline 2T-Physics to 1T field
theory\label{spin half worldine}}

There is another way to obtain 1T field theory from 2T-physics. This would
start with the worldline formalism in $d+2$ dimensions, gauge fix to $\left(
d-1\right)  +1$ dimensions to specify a shadow, and then do first quantization
of that shadow. The first quantized wavefunction is the shadow field in 1T
field theory. In this appendix we compare this procedure to the results
obtained directly from 2T field theory, and in this way illustrate the greater
power of the 2T field theory formalism.

For comparison purposes with 2T field theory, we concentrate in this appendix
on deriving the Dirac equation in curved space through first quantization of
the classical gauge fixed 2T worldline theory. \ The OSp$\left(  1|2\right)  $
gauge invariant action is \cite{spin2t}%

\begin{equation}
S=\int d\tau\left[  P\cdot\dot{X}+\frac{i}{2}\Psi\cdot\dot{\Psi}-\frac{1}%
{2}A^{ij}X_{i}\cdot X_{j}+iF^{i}X_{i}\cdot\Psi\right]  \label{worldline}%
\end{equation}
where $X_{1}^{M}\equiv X^{M}$ and $X_{2}^{M}\equiv P^{M}$ is bosonic phase
space and $\Psi^{M}$ are fermionic degrees of freedom that represent spin. The
OSp$\left(  1|2\right)  $ gauge symmetry, with bosonic and fermionic gauge
potentials $A^{ij,}F^{i},$ has 2 fermionic and three bosonic gauge parameters,
which allow us to freely choose some gauges corresponding to these parameters.
We use one of the fermionic gauge parameters to fix the fermion $\Psi
^{+^{\prime}}\left(  \tau\right)  =0$ for all $\tau$ (using the basis
$\Psi^{M}=\left(  \Psi^{+^{\prime}},\Psi^{-^{\prime}},\Psi^{m}\right)  $ in
Eq.(\ref{flat}) for the flat SO$\left(  d,2\right)  $ metric). Similarly we
use two of the bosonic gauge parameters to fix $P^{+^{\prime}}\left(
\tau\right)  =0$ and $X^{+^{\prime}}\left(  \tau\right)  =\exp\left(
\sigma\left(  X^{m}\left(  \tau\right)  \right)  \right)  $ where
$\sigma\left(  X^{m}\right)  $ is an arbitrary function of the other
coordinates. Then we solve explicitly one out of the two fermionic constraints
$X\cdot\Psi=0$ and two out of the three bosonic constraints $X^{2}=0,$ $X\cdot
P=0.$ Then $X^{M}\left(  \tau\right)  ,$ $P^{M}\left(  \tau\right)  ,$
$\Psi^{M}\left(  \tau\right)  ,$ take the following gauge fixed forms
\begin{equation}
X^{M}\left(  \tau\right)  =e^{\sigma\left(  x\left(  \tau\right)  \right)
}\left(  \overset{+^{\prime}}{1},\text{ }\overset{-^{\prime}}{\frac{1}{2}%
q^{2}\left(  x\left(  \tau\right)  \right)  },\text{ \ }\overset{m}%
{q^{m}\left(  x\left(  \tau\right)  \right)  }\right)  \label{worldlineGauge}%
\end{equation}
where we used $X^{m}=e^{\sigma\left(  x\right)  }q^{m}\left(  x\right)  $ as a
parametrization of $X^{m}$ in terms of $x^{\mu}\left(  \tau\right)  $ via $d$
arbitrary functions $q^{m}\left(  x\right)  $ (i.e. not a gauge choice), and
similarly,%
\begin{align}
P_{M}\left(  \tau\right)   &  =\left(  \overset{+^{\prime}}{0},\text{
\ \ }\overset{-^{\prime}}{q^{m}\left(  x\left(  \tau\right)  \right)
e_{m}^{\mu}\left(  x\left(  \tau\right)  \right)  p_{\mu}},\text{ \ }%
\overset{m}{e_{m}^{\mu}\left(  x\left(  \tau\right)  \right)  p_{\mu}\left(
\tau\right)  }\right) \\
\Psi^{M}\left(  \tau\right)   &  =\left(  \overset{+^{\prime}}{0},\text{
\ }\overset{-^{\prime}}{q\left(  x\left(  \tau\right)  \right)  \cdot
\zeta\left(  \tau\right)  },\text{ \ \ \ }\overset{m}{\zeta^{m}\left(
\tau\right)  }\right)
\end{align}
where $e_{m}^{\mu}\left(  x\left(  \tau\right)  \right)  $ is the inverse of
the vielbein $e_{\mu}^{m}\left(  x\right)  =e^{\sigma\left(  x\right)  }%
\frac{\partial q^{m}\left(  x\right)  }{\partial x^{\mu}}.$ The remaining so
far unsolved constraints take the form $P^{2}=g^{\mu\nu}p_{\mu}p_{\nu}=0$,
$\Psi\cdot P=\zeta^{k}e_{k}^{\mu}p_{\mu}=0.$ If we insert these forms into the
action (\ref{worldline}) we obtain the spinning particle action in a curved
background as follows.
\begin{equation}
S=\int d\tau\left[  p_{\mu}\dot{x}^{\mu}+\frac{i}{2}\zeta_{m}\dot{\zeta}%
^{m}-\frac{1}{2}A^{22}g^{\mu\nu}\left(  x\right)  p_{\mu}p_{\nu}+iF^{2}%
e_{m}^{\mu}\left(  x\right)  \zeta^{m}p_{\mu}\right]  . \label{worldl2}%
\end{equation}
Here $e_{m}^{\mu}$, $g^{\mu\nu}$ are the inverses of the emergent vielbein and
metric
\begin{equation}
e_{\mu}^{m}\left(  x\right)  =e^{\sigma\left(  x\right)  }\frac{\partial
q^{m}\left(  x\right)  }{\partial x^{\mu}},~~g_{\mu\nu}\left(  x\right)
=\eta_{mn}e_{\mu}^{m}\left(  x\right)  e_{\nu}^{n}\left(  x\right)
\end{equation}
and they are in agreement with the corresponding expressions that emerge
directly in field theoretic approach as given in Eq.(\ref{metric1}).

The fact that this is indeed the metric can be confirmed by our derivation of
the Dirac equation which is obtained in covariant first quantization by
imposing the last fermionic OSp$\left(  1|2\right)  $ constraint%
\begin{equation}
\Psi\cdot P=\zeta^{k}e_{k}^{\mu}p_{\mu}=0.
\end{equation}
We now quantize this equation, by representing the Clifford algebra among the
$\zeta^{k}$ by the SO$\left(  d-1,1\right)  $ gamma matrices $\gamma^{k}$
acting on a Dirac spinor $\psi_{\alpha}\left(  x\right)  ,$ as usual. We must
also take into account quantum ordering issues for $x,p$ in the non-linear
expression $\zeta^{k}e_{k}^{\mu}\left(  x\right)  p_{\mu}$ where $p_{\mu}$ is
replaced by a derivative.\ This ordering ambiguity leads to the addition of
some function $a_{k}\left(  x\right)  $ in the Dirac equation as shown below%
\begin{equation}
\left\{  i\gamma^{k}e_{k}^{\mu}\left(  x\right)  \partial_{\mu}+i\gamma
^{k}a_{k}\left(  x\right)  \right\}  \psi\left(  x\right)  =0
\label{Dirac with ordering}%
\end{equation}
The ambiguity $a_{k}\left(  x\right)  $ must be fixed by requiring that the
SO$\left(  d,2\right)  $ global symmetry of the worldline action
(\ref{worldline}) must be preserved at the quantum level. As it turns out (as
verified in the text) this criterion also matches with the requirement that
Eq.(\ref{Dirac with ordering}) should be compatible with the general form of
the Dirac equation in curved space%
\begin{equation}
i\gamma^{k}e_{k}^{\mu}\left(  \partial_{\mu}+\frac{1}{4}\omega_{\mu}%
^{mn}\gamma_{mn}\right)  \psi\left(  x\right)  =0.
\end{equation}
The spin connection $\omega_{\mu}^{mn}$ is generally obtained from the
vielbein $e_{\mu}^{m}$ via the well known formula in Eq.(\ref{spinConn2}). For
the vielbein of the form $e_{\mu}^{m}\left(  x\right)  =e^{\sigma\left(
x\right)  }\frac{\partial q^{m}\left(  x\right)  }{\partial x^{\mu}}$ that
emerged above, the spin connection takes the form
\begin{equation}
\omega_{\mu}^{mn}=\left(  e_{\mu}^{m}e^{n\nu}-e_{\mu}^{n}e^{m\nu}\right)
\partial_{\nu}\sigma\left(  x\right)  .
\end{equation}
Inserting this into the Dirac equation above, we can finally calculate the
spin connection term%
\begin{equation}
\frac{i}{4}\gamma^{k}e_{k}^{\mu}\omega_{\mu}^{mn}\gamma_{mn}=\frac{i}%
{2}\left(  d-1\right)  \gamma^{k}e_{k}^{\mu}\partial_{\mu}\sigma.
\end{equation}
Comparing this with $i\gamma^{k}a_{k}\left(  x\right)  $ in eq.
(\ref{Dirac with ordering}), we fix the ambiguity as%
\begin{equation}
a_{k}\left(  x\right)  =\frac{1}{2}\left(  d-1\right)  e_{k}^{\mu}%
\partial_{\mu}\sigma,
\end{equation}
and obtain the Dirac equation:%
\begin{equation}
i\gamma^{k}e_{k}^{\mu}\left(  \partial_{\mu}+\frac{1}{2}\left(  d-1\right)
\partial_{\mu}\sigma\right)  \psi\left(  x\right)  =0.
\end{equation}
This is precisely in agreement with the result obtained from 2T field theory
as seen in Eq.(\ref{result}).

Similar treatments for fields of spin-0,1 or higher would also be in agreement
with 2T field theory, because the SO$\left(  d,2\right)  $ covariant 2T field
theoretic approach automatically fixes the quantum ordering ambiguities for
any gauge of the worldline theory. In particular, we remind the reader of our
result in \cite{paper1} that when the ambiguity for the scalar field is fixed,
the resulting scalar field is the conformal scalar in a curved background
described by the action%
\begin{equation}
S\left(  \Phi\right)  _{red}=\int d^{d}x\sqrt{-g}\left(  -\frac{1}{2}g^{\mu
\nu}\partial_{\mu}\phi\partial_{\nu}\phi-\frac{d-2}{8\left(  d-1\right)
}R\phi^{2}\right)  , \label{confScalar}%
\end{equation}
where $R$ is the curvature of the space, which in our conformally flat space
is given by%
\begin{equation}
R=\left(  1-d\right)  \left[  dg^{\mu\nu}\partial_{\mu}\sigma\partial_{\nu
}\sigma+2g^{\mu\nu}\partial_{\mu}\partial_{\nu}\sigma+2e^{n\mu}\partial_{\mu
}e_{n}^{\nu}\partial_{\nu}\sigma\right]  .
\end{equation}

Another ordering ambiguity that occurs with the SO$\left(  d,2\right)  $
hidden global symmetry generators gets resolved as follows. The 2T worldline
action (\ref{worldline}) has a global SO$\left(  d,2\right)  $ symmetry with
generators%
\begin{equation}
J^{MN}=X^{M}P^{N}-X^{N}P^{M}+S^{MN},\;\;S^{MN}\equiv\frac{1}{2i}\left(
\Psi^{M}\Psi^{N}-\Psi^{N}\Psi^{M}\right)  .
\end{equation}
The gauge fixed action (\ref{worldl2}) must also have SO$\left(  d,2\right)  $
as a hidden symmetry because the generators $J^{MN}$ above are gauge invariant
since they commute with the OSp$\left(  1|2\right)  $ gauge generators
$X^{2},P^{2},X\cdot P,X\cdot\Psi,P\cdot\Psi.$ Inserting the gauge fixed forms
of $X^{M},$ $P^{M},$ $\Psi^{M}$ in the $J^{MN}$ give the correct generators of
the hidden symmetry of the worldline action (\ref{worldl2})%
\begin{align}
J^{+^{\prime}-^{\prime}}  &  =e^{\sigma\left(  x\right)  }q^{m}\left(
x\right)  e_{m}^{\mu}\left(  x\right)  p_{\mu},\;\;\;\;\;J^{+^{\prime}%
m}=e^{\sigma\left(  x\right)  }e^{m\mu}\left(  x\right)  p_{\mu}\label{jmn1}\\
J^{mn}  &  =e^{\sigma\left(  x\right)  }\left[  q^{m}\left(  x\right)
e^{n\mu}\left(  x\right)  -q^{n}\left(  x\right)  e^{m\mu}\left(  x\right)
\right]  p_{\mu}+S^{mn},\;\;S^{mn}\equiv\frac{1}{2i}\left(  \zeta^{m}\zeta
^{n}-\zeta^{n}\zeta^{m}\right) \\
J^{-^{\prime}m}  &  =e^{\sigma\left(  x\right)  }\left[  \frac{1}{2}%
q^{2}\left(  x\right)  \eta^{mn}-q^{m}\left(  x\right)  q^{n}\left(  x\right)
\right]  e_{n}^{\mu}\left(  x\right)  p_{\mu}-q_{n}\left(  x\right)  S^{mn}
\label{jmn3}%
\end{align}
These far from obvious conserved hidden symmetry charges $J^{MN}\left(
x,p\right)  $ are used with Poisson brackets to generate the phase space
transformations in the classical theory in the lower $d$ dimensions. In
particular they apply for the curved spacetimes in Table-1 marked as $\left(
cf\right)  $.
\begin{equation}
\delta_{\omega}x^{\mu}=\frac{\omega_{MN}}{2}\left\{  J^{MN},x^{\mu}\right\}
,\;\;\delta_{\omega}p_{\mu}=\frac{\omega_{MN}}{2}\left\{  J^{MN},p_{\mu
}\right\}  , ~\delta_{\omega}\zeta^{m}=\frac{\omega_{MN}}{2}\left\{
J^{MN},\zeta^{m}\right\}  . \label{poisson}%
\end{equation}
In the quantum theory the factor ordering of the operators $x^{\mu}$ and
$p_{\mu}$ must be resolved in these expressions such that they are hermitian
and correctly form the SO$\left(  d,2\right)  $ Lie algebra under quantum
commutators. This difficult problem is automatically resolved in 2T field
theory in section (\ref{globalsymm}) where the quantum ordered version of
these generators is provided.

\end{document}